\documentclass[11pt,a4paper]{article}
\usepackage{jheppub}
\usepackage{amsthm,bm,mathrsfs,stmaryrd}
\usepackage{slashed,tikz}
\usepackage{enumerate, framed}
\usepackage{youngtab,simplewick}
\usepackage{longtable}

\def\sideremark#1{\ifvmode\leavevmode\fi\vadjust{\vbox to0pt{\vss
 \hbox to 0pt{\hskip\hsize\hskip1em
 \vbox{\hsize3cm\tiny\raggedright\pretolerance10000
 \noindent #1\hfill}\hss}\vbox to8pt{\vfil}\vss}}}%

\newcommand{\be}{\begin{equation}}
\newcommand{\ee}{\end{equation}}
\newcommand{\ba}{\begin{eqnarray}}
\newcommand{\ea}{\end{eqnarray}}

\newcommand{\mt}[1]{$\mathop{#1}$}

\newcommand{\sst}{\scriptscriptstyle}

\newcommand{\nn}{\nonumber\\}
\newcommand{\eq}{&=&}

\def\d{\delta}

\def\m{\mu}

\def\s{\sigma}

\def\t{\tau}
\def\u{\upsilon}

\def\cD{{\cal D}}

\def\cL{{\cal L}}

\def\cP{{\cal P}}

\author[a]{Euihun JOUNG}
\author[a]{,\quad Luca LOPEZ}
\author[a,b]{\quad and \quad Massimo TARONNA}

\affiliation[a]{Scuola Normale Superiore and INFN\\
Piazza dei Cavalieri 7, 56126 Pisa, Italy}

\affiliation[b]{Max-Planck-Institut f\"ur Gravitationsphysik
(Albert-Einstein-Institut)\\
Am M\"uhlenberg 1, 14476 Golm, Germany}

\emailAdd{euihun.joung@sns.it}
\emailAdd{luca.lopez@sns.it}
\emailAdd{massimo.taronna@aei.mpg.de}

\title{\center Generating functions of (partially-)massless
higher-spin cubic interactions}

\abstract{
Generating functions encoding cubic interactions of (partially-)massless higher-spin fields are provided within the ambient-space formalism. They satisfy a system of higher-order partial differential equations that can be explicitly solved due to their factorized form. We find that the number of consistent couplings increases whenever the squares of the field masses take some integer values (in units of the cosmological constant) and
satisfy certain conditions. Moreover, it is shown that only the supplemental solutions can give rise to non-Abelian deformations of the gauge symmetries. The presence of these conditions on the masses is a distinctive feature of (A)dS interactions that has in general no direct counterpart in flat space.
}

\begin{document}

\maketitle

\section{Introduction}
\label{sec: intro}

Since the gravitational-interaction problem of higher-spin (HS) fields\footnote{
See e.g. \cite{Bekaert:2010hw,Sagnotti:2011qp} for recent reviews on HS field theories.
See \cite{Vasiliev:2004cp,Bekaert:2005vh} for some reviews on Vasiliev's equations,
and \cite{Giombi:2012ms,Gaberdiel:2012uj,Ammon:2012wc} for AdS/CFT-related issues.}
was overcome by turning on a cosmological constant \cite{Fradkin:1986qy,Fradkin:1987ks},
many studies have been devoted to understand HS field theories in (A)dS background. The most important result of these efforts are Vasiliev's equations \cite{Vasiliev:1988sa,Vasiliev:2003ev}, that, together with String Theory, represent the only known frameworks in which interactions of HS particles can be consistently described.
However, a deeper understanding of them
requires further investigations, especially in relation to the possibility that they be only particular members of a wider class of consistent HS theories. A preliminary step towards this goal would be to construct
the most general consistent cubic couplings.\footnote{
Many efforts have been devoted in this direction: see the references in \cite{Bekaert:2010hw}
for an exhaustive list of works,
and in particular the latest works \cite{Bekaert:2010hp,Manvelyan:2010wp,Manvelyan:2010jr,Taronna:2010qq,Polyakov:2010qs,Sagnotti:2010at,Fotopoulos:2010ay,Manvelyan:2010je,Lee:2012ku,Buchbinder:2012iz,Metsaev:2012uy,Henneaux:2012wg}
and \cite{Zinoviev:2010cr,Bekaert:2010hk,Polyakov:2011sm,Vasilev:2011xf,Polyakov:2012qj,Manvelyan:2012ww,Boulanger:2012dx}
for flat-space and (A)dS interactions involving totally-symmetric HS fields, respectively.}
Among them, only a subset is expected to be compatible with higher-order interactions \cite{Polyakov:2010sk,Taronna:2011kt,Dempster:2012vw} and lead eventually to fully non-linear theories.

Finding HS interactions in (A)dS has been the aim of our previous investigations
\cite{Joung:2011ww,Joung:2012rv,Joung:2012fv,Taronna:2012gb}, where, making use of the ambient-space formalism \cite{Fronsdal:1978vb,Biswas:2002nk}, we provided the
transverse and traceless (TT) parts of
all possible cubic interactions involving massive and massless totally-symmetric HS fields. The key point of our construction was to recast the consistency conditions for the cubic vertices into a system of linear partial differential equations (PDEs),\footnote{
See the partial result \cite{Lopez:2012pr} for the generalization of this method
to mixed-symmetry HS interactions.}
whose solutions are in one-to-one correspondence to the consistent couplings.
However, while for the case of massive and massless fields one gets relatively simple second-order PDEs, when partially-massless (PM) fields\footnote{
The PM spectrum has been first discovered for lower-spin (spin 2 and 3/2) fields  in
\cite{Deser:1983mm,Deser:1983tm,Higuchi:1986py,Higuchi:1986wu,Higuchi:1989gz,Deser:2000de,Deser:2001wx},
while its HS generalization has been considered in \cite{Deser:2001pe,Deser:2001us,
Deser:2001xr,Zinoviev:2001dt}.
The implications of PM fields to (A)dS/CFT have been discussed in \cite{Dolan:2001ih,Deser:2003gw}. In
\cite{Skvortsov:2006at,Gover:2008pt,Skvortsov:2009zu,Alkalaev:2009vm,Skvortsov:2009nv,Alkalaev:2011zv}, various formulations for the description of PM fields have been proposed.
See \cite{Deser:2006zx,Zinoviev:2006im,Francia:2008hd,Deser:2012qg} for the interactions of PM fields,
and \cite{Deser:2012qg,Joung:2012qy,deRham:2012kf,
Fasiello:2012rw,Hassan:2012gz} for their connection to conformal theories
and massive gravity.
Finally, see \cite{Dixmier:1961zz} for the representations of $SO(1,4)$\,.
} enter the interactions, due to their higher-derivative gauge transformations,
one has to solve higher-order PDEs.
In \cite{Joung:2012rv} we only provided
some examples of this kind of interactions relying on a numerical algorithm.

This paper is aimed at completing the program initiated in \cite{Joung:2012rv}
and at deriving the generating functions of cubic interactions involving PM fields.
A key point of our result
is that the number of consistent couplings
depends, in a non-trivial way, on the masses of the interacting fields.
More precisely,
whenever the squares of the masses take some integer values and
satisfy certain conditions, a new class of solutions appears.
Since the solutions---which are not of the latter type---do not
lead to deformations of the gauge symmetries,
the aforementioned conditions on the masses
represent necessary conditions for the presence of non-Abelian interactions.
After some preliminaries on the ambient-space formulation of HS interactions,
we summarize our results in
Section \ref{sec: summary}.

\subsection{Preliminaries}
\label{sec: review}

\paragraph{Ambient-space formalism}

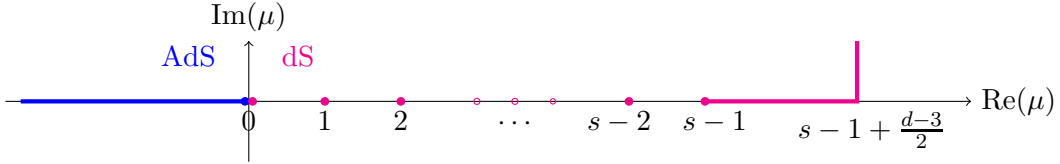
\begin{figure}[h]
\centering
\begin{tikzpicture}
\node [left] at (-.3,0.6) {\color{blue} AdS};
\node [right] at (.3,0.6) {\color{magenta} dS};
\draw [->] (-3.2,0) -- (9.5,0);
\node [right] at (9.5,0) {Re($\mu$)};
\draw [->] (0,-.8) -- (0,.8);
\node [above] at (0,.8) {Im($\mu$)};
\draw [blue,ultra thick] (-3,0) -- (0,0);
\draw [blue,fill=blue] (-.05,0) circle(.05);
\draw [magenta,fill=magenta] (.05,0) circle(.05);
\node [below] at (0,0) {$0$};
\draw [magenta,fill=magenta] (1,0) circle(.05);
\node [below] at (1,0) {$1$};
\draw [magenta,fill=magenta] (2,0) circle(.05);
\node [below] at (2,0) {$2$};
\draw [magenta] (3,0) circle(.04);
\draw [magenta] (3.5,0) circle(.04);
\node [below] at (3.5,0) {$\phantom{1}\cdots\phantom{1}$};
\draw [magenta] (4,0) circle(.035);
\draw [magenta,fill=magenta] (5,0) circle(.05);
\node [below] at (5,0) {$s-2\ \ $};
\draw [magenta,fill=magenta] (6,0) circle(.05);
\node [below] at (6,0) {$\ \ s-1$};
\draw [magenta,ultra thick] (6,0) -- (8,0) -- (8,.8);
\node [below] at (8,0) {$\ \ \ s-1+\frac{d-3}2$};
\end{tikzpicture}
\caption{Unitary values of $\mu$ (for $s>0$)\,.}
\label{fig}
\end{figure}

A way of describing totally-symmetric (A)dS tensor fields, $\varphi_{\mu_{1}\cdots\mu_{s}}$\,, is through ambient-space fields, $\Phi_{\sst M_{1}\cdots M_{s}}$\,, that are subject to
the \emph{homogeneity} and \emph{tangentiality} (HT) conditions:
\ba
	&{\rm Homogeneity}: \qquad
	&(X\cdot\partial_{X}-U\cdot\partial_{U}+2+\mu)\,\Phi(X,U)=0\,,
	\nn
	&{\rm Tangentiality}: \qquad
	&X\cdot \partial_{U}\,\Phi(X,U)=0\,,
	\label{HT}
\ea
where $\Phi(X,U)$ is the generating function of $\Phi_{\sst M_{1}\cdots M_{s}}$\,:
\be
	\Phi(X,U)=\sum_{s=0}^{\infty}\,\frac1{s!}\,
	\Phi_{\sst M_{1}\ldots M_{s}}(X)\,U^{\sst M_{1}}\cdots U^{\sst M_{s}}\,.
\ee
When
\mt{\mu=0,1, \ldots, s-1}\,, the constraints \eqref{HT} allow higher-derivative gauge
symmetries:
\be
	\delta^{\sst (0)}\,\Phi(X,U)=(U\cdot\partial_X)^{\mu+1}\,\Omega(X,U)\,,
\label{pm gt}
\ee
with gauge parameters $\Omega$ satisfying
\be
	(X\cdot\partial_X-U\cdot\partial_U-\mu)\,\Omega(X,U)=0\,,
	\qquad X\cdot\partial_U\,\Omega(X,U)=0\,.
\ee
Massless fields, \mt{\mu=0}\,, are the first members of a class of short representations where the other members, \mt{\mu=1,2,\ldots,s-1}\,, are PM fields. For other values of $\mu$,  no gauge symmetry is allowed, implying that the corresponding fields
are massive. Focussing on unitary representations,
the corresponding values of $\mu$ are constrained to
the regions shown in Figure \ref{fig}.
Let us mention that, contrary to massless fields, PM representations are unitary only in dS.

\paragraph{Cubic-interaction problem}

The most general expression for the TT parts of the cubic vertices reads
\ba
\label{cubicact1}
S^{\sst (3)} = \frac1{3!}\int_{\rm\sst (A)dS}
C(Y,Z)\ \Phi(X_1,U_1)\ \Phi(X_2,U_2)\ \Phi(X_3,U_3)\ \Big|_{^{X_i=X}_{U_i=0}}\,,
\ea
where $\int_{\rm(A)dS}$ is an integral over the codimension-one hypersurface
\mt{X^{2}=\epsilon\,L^{2}}\,(\mt{\epsilon=1} for dS and \mt{\epsilon=-1} for AdS), while
$C$ is an arbitrary function of the following parity-preserving Lorentz invariants:\footnote{
Henceforth, for brevity, we shall denote the dependence
on all six variables  by \mt{Y} and \mt{Z}.}
\be
\label{Y and Z}
Y_i=\partial_{U_i}\!\cdot\partial_{X_{i+1}}\,, \qquad
Z_i=\partial_{U_{i+1}}\!\!\cdot\partial_{U_{i-1}}
\qquad [i\simeq i+3]\,.
\ee
Assuming the $i$-th field to be (partially-)massless ((P)M)
(\emph{i.e.} \mt{\mu_{i}\in\mathbb N}),
the corresponding compatibility condition
 of the cubic vertices \eqref{cubicact1} with the gauge symmetries \eqref{pm gt} is equivalent to imposing
 \be
\label{gaugeconscond1}
\left[\,C(Y,Z)\,,
\, (U_i\!\cdot\partial_{X_i})^{\mu_i+1}\,\right]\Big|_{U_i=0}\approx 0\,.
\ee
Using Leibniz's rule, one can recast the condition \eqref{gaugeconscond1} into a higher-order PDE:
\be\label{pmdiff}
	\cL_{i}(\bar \mu_{i}-\mu_{i})\,
	\cL_{i}(\bar \mu_{i}-\mu_{i}+2)\cdots
	\cL_{i}(\bar \mu_{i}+\mu_{i})\
	C(Y,Z)=0\qquad
	\big[\,\bar\mu_{i}:=\mu_{i-1}-\mu_{i+1}\,\big]\,,
\ee
consisting in the product of the following commuting differential operators:
\be\label{L}
\cL_{i}(x):=Y_{i+1}\,\partial_{Z_{i-1}}-Y_{i-1}\,\partial_{Z_{i+1}}
+\tfrac{\hat\delta}{L}\left(Y_{i+1}\,\partial_{Y_{i+1}}-Y_{i-1}\,\partial_{Y_{i-1}}
+\tfrac12\,x \right)\,\partial_{Y_{i}}\,.
\ee
Depending on the number of (P)M fields involved in the interactions,
one can have up to three PDEs,
whose solutions
encode all possible consistent couplings.

\subsection{Summary of the results}
\label{sec: summary}

As already mentioned, a peculiar feature underlying (A)dS interactions is the appearance of
non-trivial conditions on the mass values for which the number of consistent couplings may get enhanced.
More precisely, if the $i$-th field is at one of its (P)M points \mt{\m_i\in\{0,1,\ldots,s_i-1\}}\,, the solution space of the corresponding system of PDEs may become bigger whenever the conditions
\be
\label{rmmcond1}
\mu_{i}+\mu_{i+1}-\mu_{i-1}\in 2\,\mathbb{Z}\,,
\ee
hold.  An explicit analysis shows that, while for arbitrary values of the $\mu_i$'s the solutions can be written in terms of arbitrary functions of the following operators:
\be
\tilde H_i:=\partial_{X_{i+1}}\!\cdot\partial_{X_{i-1}}\,\partial_{U_{i+1}}\!\cdot\partial_{U_{i-1}}-\partial_{U_{i-1}}\!\cdot\partial_{X_{i+1}}\,\partial_{U_{i+1}}\!\cdot\partial_{X_{i-1}}\,,
\ee
when the condition \eqref{rmmcond1} is satisfied, additional solutions involving the operator
\be
	G(Y,Z):=Y_{1}\,Z_{1}+Y_{2}\,Z_{2}+Y_{3}\,Z_{3}\,,
\ee
also appear. However, some of the $G$-couplings
can be also expressed in terms of the $\tilde H$-couplings,
so that the two kinds of solutions may have some overlap.
Let us stress that, since the $\tilde H$-couplings
are trivially gauge invariant, non-Abelian interactions are
only among those $G$-couplings which
cannot be written as $\tilde H$-couplings.
Therefore, a necessary condition for non-Abelian interactions
to be present is \eqref{rmmcond1}.
The existence of the latter mass-pattern may have also some interesting
consequences for
the interactions of one massless and two massive fields:
on the mass-pattern, indeed,
such interactions can
induce deformations of the gauge symmetries
related to non-trivial Noether currents
involving fields with \emph{different masses},
a novelty of (A)dS interactions.\footnote{
This point has been omitted in our previous work \cite{Joung:2012rv},
as,  in the analysis of the flat limit (see Appendix \ref{sec: flat}), we overlooked the singular points of the PDEs.}

Our results are summarized in the following framed paragraph.

\begin{framed}
\noindent\
\underline{\textbf{(P)M\,--\,Massive\,--\,Massive}} \bigskip \\
For arbitrary \mt{\mu_2-\mu_{3}}\,:
\be\label{H sol+}
	\tilde C =
	\sum_{\s_{1}=0}^{\m_{1}}Y_{1}^{\,\s_{1}}\,
	\tilde K^{\s_{1}}(Y_{2},Y_{3},Z_{1},\tilde H_{2},\tilde H_{3})\,.
\ee
For \mt{\mu_{1}+\mu_{2}-\mu_{3}\in 2\,\mathbb Z}\,, one also has
\ba\label{shift sol+}
	C \eq \sum_{(\t_{1},\t_{2},\t_{3})\in \mathscr L_{1}}
	Z_{1}^{\,\t_{1}}\,Z_{2}^{\,\t_{2}}\,Z^{\,\t_{3}}_{3}\
	Y_{2}^{R\left(\t_{2}+\frac{\mu_{2}-\mu_{3}-\mu_{1}}2\right)}\,
	Y_{3}^{R\left(\t_{3}+\frac{\mu_{3}-\mu_{1}-\mu_{2}}2\right)}\times \nn
	&& \hspace{60pt} \times\,
	e^{-\frac{\hat\delta}L\,\cD}\,K^{\t_{1}\t_{2}\t_{3}}(Y,G)\,\Big|_{\sst G=G(Y,Z)},
\ea
where \mt{R(x)=(x+|x|)/2} is the ramp function, $\mathcal D$ is a differential operator:
\be
	\cD:=Z_1\,\partial_{Y_2}\,\partial_{Y_3}+Z_1\,Z_2\,\partial_{Y_3}\,\partial_{G}
	+\text{cyclic}+Z_1\,Z_2\,Z_3\,\partial_G^{\,2}\,,
\ee
and $\mathscr L_{i}$ is the lattice:
\be\label{Li}
	\mathscr L_{i}:=\big\{\, (\t_{1},\t_{2},\t_{3})\in \mathbb N^{3}\ \big|\
	\t_{i+1}+\t_{i-1}\le \mu_{i}\,\big\}\,.
\ee

\noindent\
\underline{\textbf{(P)M\,--\,(P)M\,--\,Massive}} \bigskip \\
For arbitrary \mt{\mu_{3}}\,:
\be
	\tilde C =
	\sum_{\s_{1}=0}^{\mu_{1}}\sum_{\s_{2}=0}^{\m_{2}}\,
	Y_{1}^{\,\s_{1}}\,Y_{2}^{\,\s_{2}}\,
	\tilde K^{\s_{1}\s_{2}}(Y_{3},\tilde H_{1},\tilde H_{2},\tilde H_{3})\,.
\ee
For \mt{\mu_{i}+\mu_{i+1}-\mu_{i-1}\in 2\,\mathbb Z}\,, one also has
\ba\label{shift sol IS+}
	C \eq \hspace{-10pt}
	\sum_{(\t_{1},\t_{2},\t_{3})\in \mathscr L_{1}\cap\mathscr L_{2}
	\cap \mathscr L_{3}}\hspace{-10pt}
	Z_{1}^{\,\t_{1}}\,Z_{2}^{\,\t_{2}}\,Z^{\,\t_{3}}_{3}\
	Y_{1}^{R\left(\t_{1}+\frac{\mu_{1}-\mu_{2}-\mu_{3}}2\right)}\,
	Y_{2}^{R\left(\t_{2}+\frac{\mu_{2}-\mu_{3}-\mu_{1}}2\right)}\,
	Y_{3}^{R\left(\t_{3}+\frac{\mu_{3}-\mu_{1}-\mu_{2}}2\right)}\times \nn
	&& \hspace{80pt} \times\,
	e^{-\frac{\hat\delta}L\,\cD}\,K^{\t_{1}\t_{2}\t_{3}}(Y,G)\,\Big|_{\sst G=G(Y,Z)}.
\ea

\noindent\
\underline{\textbf{(P)M\,--\,(P)M\,--\,(P)M}} \bigskip \\
For arbitrary \mt{\mu_{i}}'s\,:
\be\label{Qtilde3}
	\tilde C =
	\sum_{\s_{1}=0}^{\mu_{1}}\sum_{\s_{2}=0}^{\m_{2}}\sum_{\s_{3}=0}^{\m_{3}}\,
	Y_{1}^{\,\s_{1}}\,Y_{2}^{\,\s_{2}}\,Y_{3}^{\,\s_{3}}\,
	\tilde K^{\s_{1}\s_{2}\s_{3}}(\tilde H_{1},\tilde H_{2},\tilde H_{3})\,.
\ee
For \mt{\mu_{i}+\mu_{i+1}-\mu_{i-1}\in 2\,\mathbb Z}\,, one also has
\ba\label{shift sol IS++}
	C \eq \hspace{-10pt}
	\sum_{(\t_{1},\t_{2},\t_{3})\in \mathscr L_{1}\cap\mathscr L_{2}
	\cap \mathscr L_{3}}\hspace{-10pt}
	Z_{1}^{\,\t_{1}}\,Z_{2}^{\,\t_{2}}\,Z^{\,\t_{3}}_{3}\
	Y_{1}^{R\left(\t_{1}+\frac{\mu_{1}-\mu_{2}-\mu_{3}}2\right)}\,
	Y_{2}^{R\left(\t_{2}+\frac{\mu_{2}-\mu_{3}-\mu_{1}}2\right)}\,
	Y_{3}^{R\left(\t_{3}+\frac{\mu_{3}-\mu_{1}-\mu_{2}}2\right)}\times \nn
	&& \hspace{80pt} \times\,
	e^{-\frac{\hat\delta}L\,\cD}\,K^{\t_{1}\t_{2}\t_{3}}(Y,G)\,\Big|_{\sst G=G(Y,Z)}.
\ea

\end{framed}

\subsection*{Organization of the paper}

Section \ref{sec: generality} contains some general discussions about polynomial solutions to PDEs.
In Section \ref{sec: 1 eq}, we provide the most general solutions to one PDE encoding the interactions of one (P)M field.
The solutions to the system of PDEs corresponding to general couplings
that involve  more than one (P)M field are derived in Section \ref{sec: intersection}.
Our results are discussed in Section \ref{sec: counting}.
Finally, Appendices \ref{sec: series}, \ref{sec: shift}, \ref{sec: H} and \ref{sec: flat} contain some mathematical details on the derivations of the results presented in this paper.

\section{General structure of the solutions}
\label{sec: generality}

Before solving the consistency equation \eqref{pmdiff}, let us first discuss in some detail the general idea underlying our way of organizing
the polynomial solutions. The latter are in fact the only relevant solutions for the analysis of HS cubic interactions.

\subsection{Cubic interactions as a vector space}
\label{subsec: vector space}

Restricting the attention to polynomials in $Y$ and $Z$\,, the function $C(Y,Z)$ can be expanded as
\be
	\label{polyans}
	C(Y,Z)=\sum_{\s_{i},\t_{i}\ge 0}\,C^{\t_{1}\t_{2}\t_{3}}_{\s_{1}\s_{2}\s_{3}}\,
	Z_{1}^{\,\t_{1}}\,Z_{2}^{\,\t_{2}}\,Z_{3}^{\,\t_{3}}\,
	Y_{1}^{\,\s_{1}}\,Y_{2}^{\,\s_{2}}\,Y_{3}^{\,\s_{3}}\,,
\ee
where \mt{C^{\t_{1}\t_{2}\t_{3}}_{\s_{1}\s_{2}\s_{3}}} are arbitrary coefficients. Since the operators $\cL_{\sst i}$ \eqref{L} preserve the spin degrees:
\be\label{s fix}
	s_{i}=\s_{i}+\t_{i+1}+\t_{i-1}\,,
\ee
any solution to eq.~\eqref{pmdiff} can be decomposed into solutions with fixed spins $s_{i}$\,. Moreover, being a linear PDE, for given $s_{i}$
its solution space is a finite dimensional vector space, whose elements can be labeled as:
\be\label{lin comb C}
	C^{s_{1}s_{2}s_{3}}(Y,Z)
	=\sum_{n=1}^{N}\, K_{n}\ P^{s_{1}s_{2}s_{3}}_{n}(Y,Z)
	\qquad [K_{n}\in \mathbb R]\,.
\ee
Here, $N$ is the dimension of the vector space, namely the number of \mt{s_{1}\!-\!s_{2}\!-\!s_{3}} independent couplings, whereas
\be\label{basis}
	\left\{\,P^{s_{1}s_{2}s_{3}}_{1}(Y,Z)\,,\,
	P^{s_{1}s_{2}s_{3}}_{2}(Y,Z)\,,\, \ldots\,,\,
	P^{s_{1}s_{2}s_{3}}_{N}(Y,Z)\,\right\}\,,
\ee	
is a corresponding basis. Therefore, solving eq.~\eqref{pmdiff} is tantamount to finding a set of basis vectors $P^{s_{1}s_{2}s_{3}}_{n}(Y,Z)$\,.
Let us notice that, since the spins do not enter explicitly the PDE, its solutions are spin independent. Hence, one can first determine the basis \mt{\left\{\, P_{n}(Y,Z)\, \right\}} without specifying its spin dependence, and then restrict the attention to fixed spins.

\subsection{Choice of basis}
\label{subsec: basis}

The basis \mt{\left\{\, P_{n}(Y,Z)\, \right\}} is of course not unique,
and  some choices can be  more convenient than others.
In order to clarify this point, let us first expand the basis solution $P_{n}(Y,Z)$
in powers of $\hat\delta/L$\,:
\be\label{1/L exp}
	P_{n}\big(\tfrac{\hat{\delta}}{L};Y,Z\big)=
	\sum_{k\ge 0}\,\big(\tfrac{\hat{\delta}}{L}\big)^k\,
	P^{\sst (k)}_{n}(Y,Z)\,.
\ee
In physical term, this is an expansion in the number of (ambient-space) derivatives,
so that the leading term \mt{P_{n}^{\sst (0)}} can be identified with
the highest-derivative piece of the coupling $P_{n}(Y,Z)$\,.
A convenient choice is a basis whose elements $P_{n}(Y,Z)$ have all different leading terms:
\be\label{choice}
	P^{\sst (0)}_{n}(Y,Z)\neq P^{\sst (0)}_{m}(Y,Z)\qquad
	\forall\ n\neq m\,.
\ee
Let us notice that this requirement does not fix the basis uniquely. Indeed,
denoting by $\Delta_n$ the maximum number of derivatives in $P_{n}(Y,Z)$\,,
 one can always add a lower-derivative piece $P_{m}(Y,Z)$ as
\be
	Q_{n}(Y,Z) = P_{n}(Y,Z)
	+\left(\tfrac{\hat\delta}L\right)^{\!\frac12(\Delta_n-\Delta_m)}\,P_{m}(Y,Z)\qquad
	[\Delta_n>\Delta_m]\,,
\ee
without spoiling the condition \eqref{choice}. Our strategy
in order to construct a basis of this type consists in two steps:
\begin{itemize}
\item identify all possible leading terms $P_{n}^{\sst (0)}(Y,Z)$;
\item for each $P_{n}^{\sst (0)}(Y,Z)$\,, find (if it exists) a corresponding
full solution $P_{n}(Y,Z)$\,.
\end{itemize}
As we have just explained, the solution $P_{n}(Y,Z)$ is not unique but can be always replaced by some other solutions $Q_{n}(Y,Z)$ differing by lower-derivative solutions. In the next Section we exploit this freedom in the choice of basis in order to simplify our analysis. In particular, we first construct a basis:
\be\label{basis series}
	B_{P}=\big\{\,P_{n}(Y,Z)\,\big\}\,,
\ee
proving that it spans the entire solution space of eq.~\eqref{pmdiff}, and then
we introduce another set of solutions:
\be \label{basis block}
	B_{ Q}=\big\{\,Q_{n}(Y,Z)\,\big\}\,,
\ee
whose elements satisfy \mt{Q^{\sst (0)}_{n}(Y,Z)=P^{\sst (0)}_{n}(Y,Z)}
 for all $n$\footnote{
More precisely, we shall provide linear combinations satisfying this equality.}  (so, the index $n$ can be understood as a label for different leading terms). The latter conditions ensure that $B_{\sst Q}$ be also a basis of the solution space, \emph{i.e.}, \mt{{\rm Span}(B_{\sst Q})={\rm Span}(B_{\sst P})}\,. The reason for this change of basis is that,
while $B_{\sst P}$ is convenient to prove the completeness of the solution space,
 $B_{\sst Q}$ turns out to be more suitable in constructing solutions to
more than one equation.

\section{Solutions to one equation}
\label{sec: 1 eq}

\subsection{Massless equation}
\label{subsec: massless eq}

The aim of this Section is to find the general solution to eq.~\eqref{pmdiff}\,. Since the latter consists in a product of commuting operators $\cL_{1}$\,, it is convenient to first analyze the kernel of a single operator. In particular, we start
analyzing the massless case ($\mu_1=0$), where eq.~\eqref{pmdiff} reduces to
\be\label{massless}
\cL_{1}(\bar \mu_{1})\,C(Y,Z)=0\,,
\ee
with the operator $\cL_{1}$ given by
\be
\label{1eqDelta}
\cL_{1}(\bar \mu_{1})=Y_2\,\partial_{Z_3}-Y_3\,\partial_{Z_2}
+\tfrac{\hat\delta}{L}\left(Y_2\,\partial_{Y_2}-Y_3\,\partial_{Y_3}
+\tfrac{\bar \mu_{1}}2\right)\,\partial_{Y_{1}}\,.
\ee

\subsubsection{General solution}

Eq.~\eqref{massless} can be solved as a power series in $Y_{1}$\,:
\be
\label{fullsol1e2}
C_{\s_{1}}(Y,Z)=\sum_{k=0}^{\s_{1}}\,
C^{\sst (k)}_{\s_{1}}(Y_{2},Y_{3},Z)\ \big(-\tfrac{\hat{\delta}}{L}\,\partial_{Y_1}\big)^k
\,Y_{1}^{\,\s_{1}}\,,
\ee
where $C_{\s_{1}}(Y,Z)$ denote the solutions whose highest power of $Y_{1}$ is $\s_{1}$\,. Plugging the expansion \eqref{fullsol1e2} into eq.~\eqref{massless}, one ends up with a differential recurrence relation for
$C^{\sst (k)}_{\s_{1}}$\,:
\be
\label{reccrelCnn}
\left(Y_2\,\partial_{Z_3}-Y_3\,\partial_{Z_2}\right)\,C^{\sst (k)}_{\s_{1}}(Y_{2},Y_{3},Z)
=\left(Y_2\,\partial_{Y_2}-Y_3\,\partial_{Y_3}+
\tfrac12\,\bar \mu_{1}\right)C^{\sst (k-1)}_{\s_{1}}(Y_{2},Y_{3},Z)\,,
\ee
where \mt{C^{\sst (-1)}_{\s_{1}}=0}\,. The latter can be solved iteratively starting from the leading term:
\ba
\label{C0lead}
C^{\sst (0)}_{\s_{1}}(Y_{2},Y_{3},Z) \eq K\big(Y_2,Y_{3},Z_{1},G_{1}(Y,Z)\big)\nn
	\eq \sum_{\s_{2},\s_{3},\t_{1}, \u}\,K^{\t_{1}}_{\s_{1}\s_{2}\s_{3}\u}\,Z_{1}^{\,\t_{1}}\,
	Y_{2}^{\,\s_{2}}\,Y_{3}^{\,\s_{3}}\,[G_{1}(Y,Z)]^{\u}\,,
\ea
where \mt{G_1(Y,Z):=Y_2\,Z_2+Y_3\,Z_3}\,. After summing
$C_{\sigma_{1}}$ over $\sigma_{1}$\,, the full solution reads (see Appendix \ref{sec: series} for details):
\be
\label{series sol}
	C(Y,Z)= \sum_{\s_{1},\s_{2},\s_{3},\t_{1},\u}\,K^{\t_{1}}_{\s_{1}\s_{2}\s_{3}\u}\,
	P^{\t_{1}}_{\s_{1}\s_{2}\s_{3}\u}(\bar\mu_{1};Y,Z)\,,
\ee
where the polynomial functions $P^{\t_{1}}_{\s_{1}\s_{2}\s_{3}\u}$ are defined as
\ba\label{P series basis}
	&& P^{\t_{1}}_{\s_{1}\s_{2}\s_{3}\u}(\bar\mu_{1};Y,Z) =\nn
	&&=\, \Bigg[\sum^{p+q\le\s_{1}}_{p,q\ge0}\,\frac{\left[\sigma_2+\frac{\bar\mu_{1}}2\right]_{p}
	\left[\sigma_3-\frac{\bar\mu_{1}}2\right]_q}{\left[\sigma_2+\sigma_3\right]_{p+q}}\,
	\frac{\big(-\frac{\hat\delta}L\,Z_3\,\partial_{Y_{1}}\,\partial_{Y_2}\big)^{p}}{p!}\,
	\frac{\big(-\frac{\hat\delta}L\,Z_2\,\partial_{Y_3}\,\partial_{Y_{1}}\big)^{q}}{q!}\,
	\Bigg]\times \nn
	&& \qquad \times\phantom{\big|}
	Z_{1}^{\,\t_{1}}\,Y_{1}^{\,\s_{1}}\,Y_{2}^{\,\s_{2}}\,Y_{3}^{\,\s_{3}}\,
	[G_{1}(Y,Z)]^{\u}\,,
\ea
and \mt{[a]_{n}:=a(a-1)\cdots(a-n+1)} is the descending Pochhammer symbol. Let us emphasize that, due to the presence of the terms
\mt{[\sigma_2+\sigma_3]_{p+q}} in the denominator, these functions may be ill-defined.
More precisely, potentially diverging terms appear for those values of $p$ and $q$ such that
\be\label{con pq}
	\s_{2}+\s_{3}+1\le p+q\le \s_{1}\,.
\ee
Hence, there are only two cases in which the solutions are well-defined:
\begin{itemize}
\item When
\mt{\s_{2}+\s_{3}\ge \s_{1}}\,. In this case, the problematic terms are simply absent and one gets the following set of solutions:
\be\label{BP1}
	B_{P1}(\bar\mu_{1})
	=\big\{\,P^{\t_{1}}_{\s_{1}\s_{2}\s_{3}\u}(\bar\mu_{1})\ \big|\
	\s_{2}+\s_{3}\ge \s_{1}\,\big\}\,.
\ee
\item When the condition \eqref{con pq} holds and the residue:
\ba\label{residue}
 	\left[\sigma_2+\tfrac{\bar\mu_{1}}2\right]_{p}
	\left[\sigma_3-\tfrac{\bar\mu_{1}}2\right]_{q}
	\eq (\tfrac{\bar\mu_{1}}2+\s_{2})\cdots
	(\tfrac{\bar\mu_{1}}2-\s_{3})\times\nn
	&&\times\,
	(-1)^{q}\,
	(\tfrac{\bar\mu_{1}}2+q-\s_{3}+1)\cdots
	(\tfrac{\bar\mu_{1}}2-p+\s_{2}+1)\,,
\ea
of the corresponding diverging piece
vanishes. This happens when
\be
(\tfrac{\bar\mu_{1}}2+\s_{2})\cdots
	(\tfrac{\bar\mu_{1}}2-\s_{3})=0\,,
\ee
so that the following set of solutions:
\be\label{BP2}
	B_{P2}(\bar\mu_{1})
	=\big\{\,P^{\t_{1}}_{\s_{1}\s_{2}\s_{3}\u}(\bar\mu_{1})\ \big|\
	\s_{3} \ge \tfrac{\bar\mu_{1}}2\,,\
	\tfrac{\bar\mu_{1}}2\in \mathbb Z_{\ge 0}\,;\ \
	\s_{2} \ge |\tfrac{\bar\mu_{1}}2|\,,\
	\tfrac{\bar\mu_{1}}2\in\mathbb Z_{<0}
	\,\big\}\,,
\ee
are well-defined.
\end{itemize}
Notice that $B_{P2}$
is non-empty only when $\bar\mu_{1}$ is an even integer. Moreover,
it has a non-vanishing intersection
with $B_{P1}$\,, so much so that, when \mt{\bar\mu_{1}=0}\,,
$B_{P1}$ becomes a subset of $B_{P2}$\,.
Finally, the union of the two sets\,:
\be\label{BB series}
	B_{P}(\bar\mu_{1})=
	B_{P1}(\bar\mu_{1})\cup B_{P2}(\bar\mu_{1})\,,
\ee
forms a (redundant) basis\footnote{
With a slightly abuse of notation we use the term `basis'
to denote a set of solutions which spans the entire solution space,
regardless of the fact that they be linearly independent or not.}
of the solution space of eq.~\eqref{massless}.
Comparing eq.~\eqref{BB series} to  eq.~\eqref{basis series}, one can see
 that the index $n$ in $P_{n}(Y,Z)$
 corresponds to the collective index \mt{\{\s_{1},\s_{2},\s_{3},\t_{1},\u\}}
 labeling different leading terms.

\subsubsection{Change of basis}

At this stage, we have explicitly constructed the basis \eqref{BB series} of the solution space of eq.~\eqref{massless}.
However,
although the form \eqref{P series basis} in which
 the basis solutions are written
 makes the completeness of the solution space manifest,
 it is not suitable for the analysis of more than one equation.
Therefore, in the following we construct other two sets of solutions $B_{\tilde Q}$ and
 $B_{Q}$\,, and, analyzing their leading terms, we prove that their
 union spans the same space as
$B_{ P}$\,.

\paragraph{$\tilde Q$ solutions}
Let us notice that the operators:\footnote{
These operators can be obtained as deformations of their flat-space counterparts (see Appendix \ref{sec: flat}).}
\be
\label{tildeH}
\tilde H_i:=\partial_{X_{i+1}}\!\cdot\partial_{X_{i-1}}\,\partial_{U_{i+1}}\!\cdot\partial_{U_{i-1}}-\partial_{U_{i-1}}\!\cdot\partial_{X_{i+1}}\,\partial_{U_{i+1}}\!\cdot\partial_{X_{i-1}}\,,
\ee
commute with the gradient operators:
\be
	[\,\tilde H_{i}\,,\,U_{j}\cdot\partial_{X_{j}}\,]=0\,,
\ee
without relying on the on-shell conditions.	
As a consequence, one can easily construct couplings of the form:
\be\label{1mNoInt2}
	\tilde K(Y_2,Y_3,Z_1,\tilde{H}_2,\tilde{H}_3)\,,
\ee
that are invariant under the gauge transformations associated to the first field.
These couplings are not written as functions of $Y$ and $Z$\,, but they can be always
brought to that form by performing the integrations by parts of all the total-derivative terms present in $\tilde H_{2}$ and $\tilde H_{3}$\,.
Hence, one can consider the following set:
\be
\label{BQ tilde}
	B_{\tilde Q}(\bar\mu_{1})=
	\big\{\,\tilde Q^{\t_{1}}_{\s_{2}\s_{3}h_{2}h_{3}}(\bar\mu_{1})\,\big\}\,,
\ee
whose elements are given by
\be
\label{tilde Q fn}
	\tilde Q^{\t_{1}}_{\s_{2}\s_{3}h_{2}h_{3}}(\bar\mu_{1};Y,Z)
	\simeq
	Z_1^{\,\t_{1}}\,Y_2^{\,\s_2}\,Y_3^{\,\s_3}\,\tilde{H}_2^{\,h_2}\,\tilde{H}_3^{\,h_3}\,,
\ee
where $\simeq$ means equivalence modulo integrations by parts.
Although finding the exact form of $\tilde Q^{\t_{1}}_{\s_{2}\s_{3}h_{2}h_{3}}$
requires an involved analysis (see Appendix~\ref{sec: H} for details), for our purpose
it is sufficient to identify the corresponding leading terms:
\be
\label{Cbblead}
\tilde Q^{\t_{1}}_{\s_{2}\s_{3}h_{2}h_{3}}(\bar\mu_{1};Y,Z)=
Z_1^{\,\t_{1}}\,Y_{1}^{h_{2}+h_{3}}\,
Y_2^{\s_2+h_{3}}\,Y_3^{\s_3+h_{2}}+
\mathcal O\big(\tfrac{\hat\delta}L\big)\,.
\ee
At this point, it is straightforward to check that these leading terms exactly reproduce the ones of $P^{\t_{1}}_{\s_{1}\s_{2}\s_{3}\u}$
 in $B_{P1}$ \eqref{BP1} for \mt{\u=0}.
The dependence  on $G_1$ can be recovered by taking
proper combinations of them. For instance, the \mt{3\!-\!2\!-\!3} coupling starting with the leading term \mt{Y_1\,Y_3\,G_1^2}  is given by
\ba\label{ex H int}
	 &&\left[\tfrac{\bar\mu_{1}-2}2\,\big(Y_2\,\tilde{H}_2\big)^{2}\,
	 -\bar\mu_{1}\,\big(Y_2\,\tilde{H}_2\big)
	 \big(Y_{3}\,\tilde H_{3}\big)
	 +\tfrac{\bar\mu_{1}+2}2\,\big(Y_3\,\tilde{H}_3\big)^{2}\right]
	 \tilde H_{2}\nn
	&&\simeq \big(\tfrac{\hat\delta}{L}\big)^2\,
	\tfrac{\bar\mu_{1}-2}2\,\tfrac{\bar\mu_{1}}2\,\tfrac{\bar\mu_{1}+2}2
	\left[Y_1\,Y_3\,G_1^{\,2}
	+\tfrac{\hat\delta}{L}\,Z_2\,G_{1}
	\left(\tfrac{\bar\mu_{1}-2}2\,Y_2\,Z_2
	+\tfrac{\bar\mu_{1}-6}2\,Y_3\,Z_3\right)\right].
\ea
As one can see from this example, depending on the values of $\bar\mu_{1}$\,,
the relation between the elements of the two basis $B_{\tilde Q}(\bar\mu_{1})$
and $B_{P1}(\bar\mu_{1})$ can be singular.
However,
the vanishing coefficient (\emph{i.e.}, \mt{(\bar\mu_{1}+2)\,\bar\mu_{1}\,(\bar\mu_{1}-2)} in \eqref{ex H int})
is overall so that one can always normalize the leading term.
See Appendix \ref{sec: H} for the general case.
Since all the leading terms of $B_{P1}(\bar\mu_{1})$
can be reproduced by $\tilde Q^{\t_{1}}_{\s_{2}\s_{3}h_{2}h_{3}}(\bar\mu_{1})$\,,
and $B_{P1}(\bar\mu_1)$ covers the whole solution space when \eqref{rmmcond1} is not satisfied,
one can conclude that\footnote{By continuity of the solution spaces $B_{P1}(\bar\m_1)$ and $B_{\tilde Q}(\bar\mu_{1})$ in $\bar\m_1$\,,
this statement holds also when the condition \eqref{rmmcond1} is satisfied.}
\be
	{\rm Span}\big(B_{\tilde Q}(\bar\mu_{1})\big)=
	{\rm Span}\big(B_{P1}(\bar\mu_{1})\big)\,.
\ee

\paragraph{$Q$ solutions}

Let us notice that, when the constant $\bar\mu_{1}/2$ in eq.~\eqref{massless} is an integer number, it can be removed
factoring terms of the form $Y_{2}^{-\bar\mu_{1}/2}$ or $Y_{3}^{\bar\mu_{1}/2}$\,.
Hence, using the ramp function
\be
R(x):=(|x|+x)/2\,,
\ee
one can construct the following solutions:
 \be
	\label{sol1eqInt}
	C(Y,Z)=Y_{2}^{R\left(-\frac{\bar\mu_{1}}2\right)}\,
	Y_{3}^{R\left(\frac{\bar\mu_{1}}2\right)}\,
	e^{-\frac{\hat{\delta}}{L}\,\cD}\,
	K(Y_1,Y_2,Y_3,Z_1,G)\,\Big|_{\sst G=G(Y,Z)}
	\qquad [\,{\rm if}\ \tfrac{\bar\mu_{1}}2\in \mathbb Z\,]\,,
\ee
with
\be\label{G def}
	G(Y,Z):=Y_{1}\,Z_{1}+Y_{2}\,Z_{2}+Y_{3}\,Z_{3}\,.
\ee
Here $e^{-\frac{\hat{\delta}}{L}\,\cD}\,K$ are the general solutions to
the massless equation
\mt{\cL_{1}(0)\,C=0}\,, where the operator $\cD$ is defined as
\be
	\cD:=Z_1\,\partial_{Y_2}\,\partial_{Y_3}+Z_1\,Z_2\,\partial_{Y_3}\,\partial_{G}
	+\text{cyclic}+Z_1\,Z_2\,Z_3\,\partial_G^{\,2}\,.
\ee
The solutions \eqref{sol1eqInt} can be decomposed in terms of
the following functions:
\ba
	&& Q^{\t_{1}}_{\s_{1}\s_{2}\s_{3}\u}(\bar\mu_{1};Y,Z)
	:=Z_{1}^{\,\t_{1}}\,
	Y_{2}^{R\left(-\frac{\bar\mu_{1}}2\right)}\,Y_{3}^{R\left(\frac{\bar\mu_{1}}2\right)}\,
	e^{-\frac{\hat{\delta}}{L}\,\cD}\,Y_{1}^{\s_{1}}\,
	Y_{2}^{\s_{2}}\,Y_{3}^{\s_{3}}\,G^{\u}\,\Big|_{\sst G=G(Y,Z)}\nn
	&&\quad=\,Z_{1}^{\,\t_{1}}\,
	Y_{1}^{\s_{1}}\,Y_{2}^{\s_{2}+R\left(-\frac{\bar\mu_{1}}2\right)}\,
	Y_{3}^{\s_{3}+R\left(\frac{\bar\mu_{1}}2\right)}\,
	[G_{1}(Y,Z)+Y_{1}\,Z_{1}]^{\u}+
	\mathcal O\big(\tfrac{\hat\delta}L\big)\,,
\ea
whose leading terms coincide with those of $P^{\t_{1}}_{\s_{1}\s_{2}\s_{3}\u}$ in $B_{P2}$ \eqref{BP2}.
As a consequence, the set:
\be
\label{BQ}
	B_{Q}(\bar\mu_{1})
	=\big\{\,Q^{\t_{1}}_{\s_{1}\s_{2}\s_{3} \u}(\bar\mu_{1})\,\big\}\,,
\ee
together with $B_{\tilde Q}(\bar\mu_{1})$ span the entire solution space\,:
\be\label{QP2}
	{\rm Span}\big(B_{\tilde Q}(\bar\mu_{1})\cup B_{Q}(\bar\mu_{1})\big)=
	{\rm Span}\big(B_{P1}(\bar\mu_{1})\cup B_{P2}(\bar\mu_{1})\big)\,.
\ee

\subsection{Partially-massless equation}
\label{subsec: PM eq}

Let us now move to the case in which one PM field (\mt{
\mu_{1}\in \mathbb N}) is involved in the interactions. Then, the corresponding cubic vertices have to satisfy the PDE:
\be\label{pmless}
	\cL_{1}(\bar \mu_{1}-\mu_{1})\,\cL_{1}(\bar \mu_{1}-\mu_{1}+2)
	\cdots \cL_{1}(\bar \mu_{1}+\mu_{1})\,C(Y,Z)=0\,.
\ee

\subsubsection{General solutions}

The general solutions to the above equation
can be decomposed in terms of
the functions $C_{k}$'s satisfying
\be
	\cL_{1}(\bar\mu_{1}-\mu_{1}+2\,k)\,C_{k}=0\,,
\ee
with \mt{k=0,1,\ldots,\mu_{1}}.
The $C_{k}$'s are given by \eqref{series sol}
where  \mt{\bar\mu_{1}} is replaced by \mt{\bar\mu_{1}-\mu_{1}+2\,k}\,,
hence the following set of functions:
\be\label{formal pm set}
	\big\{\,P^{\t_{1}}_{\s_{1}\s_{2}\s_{3}\u}(\bar\mu_{1}-\mu_{1})\,\big\}
	\,\cup\,
	\big\{\,
	P^{\t_{1}}_{\s_{1}\s_{2}\s_{3}\u}(\bar\mu_{1}-\mu_{1}+2)\,\big\}
	\,\cup\,\cdots\,\cup\,
	\big\{\,
	P^{\t_{1}}_{\s_{1}\s_{2}\s_{3}\u}(\bar\mu_{1}+\mu_{1})\,\big\}\,,
\ee
forms a complete basis for the solution space of eq.~\eqref{pmless}.
However, this decomposition is possible only when $\s_{2}$ and $\s_{3}$
are regarded as real (non-integer) numbers.
Indeed, as in the massless case, the set \eqref{formal pm set}
contains solutions that are ill-defined.
Therefore, out of them, one needs to select only the ones that are well-behaved
as $\s_{2}$ and $\s_{3}$ approach some integer numbers.
In the massless case, for any given leading term there is only one function in the set, so that the selection simply amounts in examining whether each function $P^{\t_{1}}_{\s_{1}\s_{2}\s_{3}\u}(\bar\mu_{1})$ is well-defined or not.
On the other hand,  in the PM case one has \mt{\mu_{1}+1} solutions
\mt{P^{\t_{1}}_{\s_{1}\s_{2}\s_{3}\u}(\bar\mu_{1}-\mu_{1}+2\,k)}
with the same leading term, so one has to analyze all possible linear combinations of them.

\paragraph{$P1$ solutions}
Let us consider the first divergent term
in the series expansion  of
$P^{\t_{1}}_{\s_{1}\s_{2}\s_{3}\u}(x)$ \eqref{P series basis}.
It arises for \mt{p+q=\s_{2}+\s_{3}+1} and its residue is \mt{(\tfrac{x}2+\s_{2})\cdots
(\tfrac{x}2-\s_{3})\,(-1)^{q}} (see eq.~\eqref{residue}).
Therefore, it can be cancelled by taking the following
linear combination:
\be\label{comb 2}
	\big(\tfrac{x}2+1+\s_{2}\big)\,
	P^{\t_{1}}_{\s_{1}\s_{2}\s_{3}\u}(x)
	-\big(\tfrac{x}2-\s_{3}\big)\,
	P^{\t_{1}}_{\s_{1}\s_{2}\s_{3}\u}(x+2)\,.
\ee
Although one can get rid of the first divergent term,
there is no way to eliminate the remaining divergent pieces that are
still present for \mt{p+q\ge \s_{2}+\s_{3}+2}\,.
Hence, the combination \eqref{comb 2}
is well-defined only when the order of the polynomial
is low enough in order to prevent those divergent terms from showing up,
that is, when \mt{\s_{2}+\s_{3}+1\ge \s_{1}}.
Let us notice that this condition on the leading term is very similar to the one
obtained in the massless case \eqref{BP1}, but for the shift by one on the left-hand side
of the inequality.
In general, taking the following linear combination:
\ba\label{P[n]}
	&&P^{[n]\,\t_{1}}_{\s_{1}\s_{2}\s_{3}\u}(x):=\nn
	&&=\,
	\sum_{k=0}^{n}\,(-1)^{k}\,\binom{n}{k}
	\left[\tfrac{x}2+n+\s_{2}\right]_{n-k}
	\left(\tfrac{x}2-\s_{3}\right)_{k}\,
	P^{\t_{1}}_{\s_{1}\s_{2}\s_{3}\u}(x+2k)\,,
\ea
one can cancel the first $n$ divergent terms. Once
again, as  these functions contain divergent terms for \mt{p+q\ge \s_{2}+\s_{3}+n+1},
only the ones belonging to the set:
\be\label{BP1 n}
	B^{[n]}_{P1}(x)=\big\{\,
	 P^{[n]\,\t_{1}}_{\s_{1}\s_{2}\s_{3}\u}(x)\ |\
	 \s_{2}+\s_{3}+n\ge \s_{1}\ge n\,\big\}\,,
\ee
are well-defined. Here, the lower bound on $\s_{1}$ has been
introduced since the solutions $P^{[n]\,\t_{1}}_{\s_{1}\s_{2}\s_{3}\u}$
with \mt{\s_{1}<n}---not having a
sufficient number of terms---are simply linear combinations
of \mt{P^{[0]\,\t_{1}}_{\s_{1}\s_{2}\s_{3}\u}\,,\,
P^{[1]\,\t_{1}}_{\s_{1}\s_{2}\s_{3}\u}\,,\,
\ldots\,,\,P^{[n-1]\,\t_{1}}_{\s_{1}\s_{2}\s_{3}\u}}\,.
To sum up, starting from the formal basis \eqref{formal pm set},
one can extract the following set of well-defined solutions:
\be\label{pm P1}
	{\bm B}_{P1}(\bar\mu_{1},\mu_{1}):=B^{[0]}_{P1}(\bar\mu_{1}-\mu_{1})\,\cup\,
	B^{[1]}_{P1}(\bar\mu_{1}-\mu_{1})\,\cup\,\cdots\,\cup\,
	B^{[\mu_{1}]}_{P1}(\bar\mu_{1}-\mu_{1})\,.
\ee
	
\paragraph{$P2$ solutions}	

On the other hand, precisely as in the massless case \eqref{BP2}, the divergent terms are harmless when the corresponding residues vanish.
Since everything is well-defined in this case,
the corresponding solution space is spanned by the union of the massless sets \eqref{BP2}:
\be\label{pm P2}
	B_{P2}(\bar\mu_{1}-\mu_{1})\,\cup\,
	B_{P2}(\bar\mu_{1}-\mu_{1}+2)\,\cup\,\cdots\,
	\cup\,B_{P2}(\bar\mu_{1}+\mu_{1})\,.
\ee
However, for the succeeding analysis it proves convenient to change basis.
For this purpose, let us first introduce the new basis functions:
\be\label{P2PM}
	P^{\{n\}\t_{1}}_{\s_{1}\s_{2}\s_{3}\u}(x):=
	\sum_{k=0}^{|n|}\,(-1)^{k}\,\binom{|n|}{k}\,
	P^{\t_{1}}_{\s_{1}\s_{2}\s_{3}\u}(x+{\rm sgn}(n)\,2k)\,,
\ee
which are  linear combinations (with binomial coefficients)
of \mt{|n|+1} functions with consecutive arguments
starting from $x$\,.
The functions \eqref{P2PM} are well-defined for $n\ge0$ if \mt{\s_{3}\ge x/2+n}
and for $n<0$ if \mt{\s_{2}\ge x/2-n}\,.
As we will see in the next Section, these linear combinations
allow the compensation of the first $|n|$ terms in the $Y_{1}$-expansion,
making the link to the PM counterpart of the $Q$ basis function more straightforward.
In terms of these functions,  one can define the following sets:
\be\label{B [n]}
	 B^{[n]}_{P2}(\bar\mu_{1},\mu_{1})
	=\left\{
	\begin{array}{cc}
	\Big\{\,P^{\{n-R(\frac{\bar\mu_{1}-\mu_{1}}2)\}\t_{1}}_{\s_{1}\s_{2}\s_{3}\u}
	\big(R(\bar\mu_{1}-\mu_{1})\big)\ \big|
	\ \s_{3}\ge n\, \Big\}
	\ & [n\ge0]  \medskip \\
	\Big\{\,P^{\{n+R(-\frac{\bar\mu_{1}+\mu_{1}}2)\}\t_{1}}_{\s_{1}\s_{2}\s_{3}\u}
	\big(-R(-\bar\mu_{1}-\mu_{1})\big)\ \big|
	\ \s_{2}\ge |n|\, \Big\}
	\ & [n<0]
	\end{array}\right.,
\ee
such that their union:
\be\label{pm B conv}
	{\bm B}_{P2}(\bar\mu_{1},\mu_{1}):=
	B_{P2}^{[\frac{\bar\mu_{1}-\mu_{1}}2]}(\bar\mu_{1},\mu_{1})
	\, \cup\, B_{P2}^{[\frac{\bar\mu_{1}-\mu_{1}+2}2]}(\bar\mu_{1},\mu_{1})\,
	\cup\,\cdots\,\cup\,
	B_{P2}^{[\frac{\bar\mu_{1}+\mu_{1}}2]}(\bar\mu_{1},\mu_{1})\,,
\ee
spans the same space as \eqref{pm P2}.
To be concrete, let us consider the \mt{\mu_{1}=2} case,
where one can distinguish between three different subcases:
$\bar\mu_{1}\ge2$\,, $\bar\mu_{1}\le -2$ and $\bar\mu_{1}=0$\,.
If $\bar\mu_{1}\ge 2$\,, then one gets
\ba
	B_{P2}^{[\frac{\bar\mu_{1}-2}2]}\eq\left\{
	P^{\{0\}\t_{1}}_{\s_{1}\s_{2}\s_{3}\u}(\bar\mu_{1}-2)=
	P^{\t_{1}}_{\s_{1}\s_{2}\s_{3}\u}(\bar\mu_{1}-2)
	\,\Big|\,\sigma_{3}\ge \tfrac{\bar\mu_{1}-2}2 \right\}\,,\nn
	B_{P2}^{[\frac{\bar\mu_{1}}2]}\eq\left\{
	P^{\{ 1\}\t_{1}}_{\s_{1}\s_{2}\s_{3}\u}(\bar\mu_{1}-2)
	=P^{\t_{1}}_{\s_{1}\s_{2}\s_{3}\u}(\bar\mu_{1}-2)
	-P^{\t_{1}}_{\s_{1}\s_{2}\s_{3}\u}(\bar\mu_{1})
	\,\Big|\,\sigma_{3}\ge \tfrac{\bar\mu_{1}}2\right\}\,,\nn
	B_{P2}^{[\frac{\bar\mu_{1}+2}2]}\eq\Big\{P^{\{2\}\t_{1}}_{\s_{1}\s_{2}\s_{3}\u}(\bar\mu_{1}-2)=\\
	&&\hspace{10pt}
	=\,P^{\t_{1}}_{\s_{1}\s_{2}\s_{3}\u}(\bar\mu_{1}-2)
	-2\,P^{\t_{1}}_{\s_{1}\s_{2}\s_{3}\u}(\bar\mu_{1})
	+P^{\t_{1}}_{\s_{1}\s_{2}\s_{3}\u}(\bar\mu_{1}+2)
	\,\Big|\,\sigma_{3}\ge \tfrac{\bar\mu_{1}+2}2\Big\}\,,
	\nonumber
\ea
while, if $\bar\mu_{1}\le -2$ the corresponding sets are given by
\ba
	B_{P2}^{[\frac{\bar\mu_{1}-2}2]}\eq
	\Big\{P^{\{-2\}\t_{1}}_{\s_{1}\s_{2}\s_{3}\u}(\bar\mu_{1}+2)=\nn
	&&\hspace{10pt}
	=\,P^{\t_{1}}_{\s_{1}\s_{2}\s_{3}\u}(\bar\mu_{1}+2)
	-2\,P^{\t_{1}}_{\s_{1}\s_{2}\s_{3}\u}(\bar\mu_{1})
	+P^{\t_{1}}_{\s_{1}\s_{2}\s_{3}\u}(\bar\mu_{1}-2)
	\,\Big|\,\sigma_{2}\ge |\tfrac{\bar\mu_{1}-2}2|\Big\}\,,\nn
	B_{P2}^{[\frac{\bar\mu_{1}}2]}\eq\left\{
	P^{\{ -1\}\t_{1}}_{\s_{1}\s_{2}\s_{3}\u}(\bar\mu_{1}+2)
	=P^{\t_{1}}_{\s_{1}\s_{2}\s_{3}\u}(\bar\mu_{1}+2)
	-P^{\t_{1}}_{\s_{1}\s_{2}\s_{3}\u}(\bar\mu_{1})
	\,\Big|\,\sigma_{2}\ge |\tfrac{\bar\mu_{1}}2|\right\}\,,\nn
	B_{P2}^{[\frac{\bar\mu_{1}+2}2]}\eq\left\{
	P^{\{0\}\t_{1}}_{\s_{1}\s_{2}\s_{3}\u}(\bar\mu_{1}+2)=
	P^{\t_{1}}_{\s_{1}\s_{2}\s_{3}\u}(\bar\mu_{1}+2)
	\,\Big|\,\sigma_{2}\ge |\tfrac{\bar\mu_{1}+2}2| \right\}\,.
\ea
Finally, in the last case \mt{\bar\mu_{1}=0}, one ends up with
\be
	B_{P2}^{[-1]}=\left\{P^{\{-1\}\t_{1}}_{\s_{1}\s_{2}\s_{3}\u}(0)\,
	\Big|\,\s_{2}\ge 1\right\}\,,
	\quad
	B_{P2}^{[0]}=\left\{P^{\{0\}\t_{1}}_{\s_{1}\s_{2}\s_{3}\u}(0)\right\}\,,
	\quad
	B_{P2}^{[1]}=\left\{P^{\{1\}\t_{1}}_{\s_{1}\s_{2}\s_{3}\u}(0)\,
	\Big|\,\s_{3}\ge 1\right\}\,.
\ee
As one can see from the above example,
the set \eqref{B [n]} is chosen such that its elements are
linear combinations involving always the $P^{\t_{1}}_{\s_{1}\s_{2}\s_{3}\u}(x)$
with \mt{x={\rm min}\{|\bar\mu_{1}-\mu_{1}|,\ldots,|\bar\mu_{1}+\mu_{1}|\}}.

All in all, the solution space of the partially-massless equation \eqref{pmless}
is spanned by the union of
the two sets \eqref{pm P1} and \eqref{pm B conv}.

\subsubsection{Change of basis}

After having generalized the sets $B_{P1}$ and $B_{P2}$ to the sets \eqref{pm P1} and \eqref{pm B conv}, we now aim
at finding the analogue of the convenient basis
$B_{\tilde Q}$ \eqref{BQ tilde} and
$B_{Q}$ \eqref{BQ} for the partially-massless case.

\paragraph{$\tilde Q$ solutions}

The first observation is that the identity:
\ba
	&&\big[\,Y_{1}^{\,n}\,\tilde K(Y_{2},Y_{3},Z_{1},\tilde H_{2},\tilde H_{3})\,,\,(U_{1}\cdot\partial_{X_{1}})^{\mu_{1}+1}\,\big]
	=0 \qquad
	[\,n=0,1,\ldots,\mu_{1}\,]\,,
\ea
holds because the operator \mt{\tilde K(Y_{2},Y_{3},Z_{1},\tilde H_{2},\tilde H_{3})}
commutes with \mt{U_{1}\cdot\partial_{X_{1}}}
(without on-shell conditions).
Thus, let us consider the functions:
\be
	\tilde Q^{\t_{1}}_{[n]\s_{2}\s_{3}h_{2}h_{3}}(Y,Z)\simeq
	Z_{1}^{\t_{1}}\,Y^{\,n}_{1}\,
	Y_2^{\,\s_2}\,Y_3^{\,\s_3}\,\tilde{H}_2^{\,h_2}\,\tilde{H}_3^{\,h_3}\,,
\ee
that are of the same form as \eqref{tilde Q fn} but for an additional factor of $Y_{1}^{n}$\,.
In the previous sections, we have shown that, for \mt{n=0}, the sets $B_{\tilde Q}$ \eqref{BQ tilde}
 reproduce all the leading terms with \mt{\s_{2}+\s_{3}\ge \s_{1}} in $B_{P1}$\,.
For $n\ge1$, it is still manifest that the following set:
\be \label{BQ [n]}
	B^{[n]}_{\tilde Q}(\bar\mu_{1},\mu_{1})=
	\big\{\,\tilde Q^{\t_{1}}_{[n]\s_{2}\s_{3}h_{2}h_{3}}\,\big\}\,,
\ee
gives leading terms with \mt{\s_{2}+\s_{3}+n\ge \s_{1}\ge n} and $\u=0$\,,
 reproducing the same bound \eqref{BP1 n} on $\s_{1}$\,.
However, for a given leading term, depending on the values of the $\s_{i}$'s,
there can be more than one solutions---more than one values of $n$
can satisfy the inequality---in both ($P1$ and $\tilde Q$ solutions) sides.
In order to handle this subtlety as well as the $\u\ge1$ case,
one can consider once again a change of basis
in such a way to remove the degeneracy between the leading terms.
Concerning the $P1$ solutions, this can be done by taking particular linear combinations
of the $P^{[n]\,\t_{1}}_{\s_{1}\s_{2}\s_{3}\u}$'s,
with \mt{n=\mu_{1},\mu_{1}-1,\ldots, \mu_{1}-m}\,, that
give rise to the cancelation of the first $m$ leading terms.
A similar analysis can be also done for the $\tilde Q$ solutions
after integrating by parts the total-derivative terms present in $\tilde{H}_2$ and $\tilde{H}_3$  (see Appendix \ref{sec: H} for more details).
In the end, one can see that all the leading terms of the $P1$ solutions
in the set \eqref{BP1 n}
can be reproduced by the $\tilde Q$ solutions in the sets \eqref{BQ [n]}.
Hence, exploiting the completeness of the $P1$ solutions when
the condition \eqref{rmmcond1} is not satisfied,
one can conclude that
\be
	{\rm Span}\big[ {\bm B}_{\tilde Q}(\bar \mu_{1},\mu_{1})\big]
	={\rm Span}\big[{\bm B}_{P1}(\bar \mu_{1},\mu_{1})\big]\,,
\ee
where
\be
	{\bm B}_{\tilde Q}(\bar \mu_{1},\mu_{1}):=
	\bigcup_{n=0}^{\mu_{1}}B^{[n]}_{\tilde Q}(\bar\mu_{1},\mu_{1})\,.
\ee

\paragraph{$Q$ solutions}

Here, we search for a generalization of the solutions \eqref{sol1eqInt}
to the partially-massless case.\footnote{
Because of \eqref{QP2} and \eqref{pm P2}, one is naturally led to consider the basis:
\be
	B_{Q}(\bar\mu_{1}-\mu_{1})\,\cup\,
	B_{Q}(\bar\mu_{1}-\mu_{1}+2)\,\cup\,\cdots\,
	\cup\,B_{Q}(\bar\mu_{1}+\mu_{1})\,.
\ee
However, the latter is not suitable to study the solution space of more than one PDE.}
When \mt{\mu_{1}=1}, due to the identity:
\ba\label{shift id}
	&&\mathcal L_{1}(\bar\mu_{1}-1)\,\mathcal L_{1}(\bar\mu_{1}+1)\,
	Z_{2}
	=[Z_{2}\,\mathcal L_{1}(\bar\mu_{1}+1)-2\,Y_{3}]\,
	\mathcal L_{1}(\bar\mu_{1}-1)\,,\nn
	&&\mathcal L_{1}(\bar\mu_{1}-1)\,\mathcal L_{1}(\bar\mu_{1}+1)\,
	Z_{3}
	=[Z_{3}\,\mathcal L_{1}(\bar\mu_{1}-1)+2\,Y_{2}]\,
	\mathcal L_{1}(\bar\mu_{1}+1)\,,
\ea
one can easily check that the functions:
\ba
	C(Y,Z)\eq Z_{2}\ Y_{2}^{R\left(-\frac{\bar\mu_{1}-1}2\right)}\,
	Y_{3}^{R\left(\frac{\bar\mu_{1}-1}2\right)}\,
	e^{-\frac{\hat{\delta}}{L}\,\cD}\,
	K_{1}(Y_1,Y_2,Y_3,Z_1,G)\nn
	&& +\ Z_{3}\ Y_{2}^{R\left(-\frac{\bar\mu_{1}+1}2\right)}\,
	Y_{3}^{R\left(\frac{\bar\mu_{1}+1}2\right)}\,
	e^{-\frac{\hat{\delta}}{L}\,\cD}\,
	K_{2}(Y_1,Y_2,Y_3,Z_1,G)\,\Big|_{\sst G=G(Y,Z)}\,,
\ea
solve the PDE
\mt{\mathcal L_{1}(\bar\mu_{1}-1)\,\mathcal L_{1}(\bar\mu_{1}+1)\,C(Y,Z)=0}\,.
In general, one can consider the set:
\be\label{BQ pm}
	B^{[\t_{2},\t_{3}]}_{Q}(\bar\mu_{1},\mu_{1})=\big\{\,
	Q^{[\t_{2},\t_{3}]\t_{1}}_{\s_{1}\s_{2}\s_{3} \u}\,\big\}\,,
\ee
whose elements are given by
\ba\label{Q Z pm}
	Q^{[\t_{2},\t_{3}]\t_{1}}_{\s_{1}\s_{2}\s_{3} \u}(Y,Z)
	\eq Z_{1}^{\,\t_{1}}\,Z_{2}^{\,\t_{2}}\,Z^{\,\t_{3}}_{3}\,
	Y_{2}^{R\left(\t_{2}-\frac{\bar\mu_{1}+\mu_{1}}2\right)}\,
	Y_{3}^{R\left(\t_{3}+\frac{\bar \mu_{1}-\mu_{1}}2\right)}\times\nn
	&&\times\,e^{-\frac{\hat\delta}L\,\cD}\,
	Y_{1}^{\,\s_{1}}\,Y_{2}^{\,\s_{2}}\,Y_{3}^{\,\s_{3}}\,G^{\,\u}\,\Big|_{\sst G=G(Y,Z)}\,,
\ea
with \mt{\t_{2}+\t_{3}\le \mu_{1}} (see Appendix \ref{sec: shift} for details).
The proof that the union of these sets together the $\tilde Q$ solutions span the entire
solution space relies once again on the leading term analysis.
Using the identity \eqref{P lin comb}, one can show that
\ba\label{Pn(x)}
	&& P^{\{n\}\t_{1}}_{\s_{1}\s_{2}\s_{3}n}(x)=
	\big(\tfrac{\hat\delta}L\big)^{|n|}\left[\left(Z_{2}\,\partial_{Y_{3}}-Z_{3}\,	
	\partial_{Y_{2}}\right)^{|n|}
	\,Z_{1}^{\,\t_{1}}\,Y_{1}^{\,\s_{1}}\,Y_{2}^{\,\s_{2}}\,Y_{3}^{\,\s_{3}}\,
	[G_{1}(Y,Z)]^{\u}+\mathcal O\big(\tfrac{\hat\delta}L\big)\right] \nn
	&&=\,\big(\tfrac{\hat\delta}L\big)^{|n|}
	\!\!\!\sum_{\t_{2}+\t_{3}=|n|}\!\!
	(-1)^{\t_3}\,\binom{|n|}{\t_{3}}\,[\s_{2}]_{\t_{3}}\,[\s_{3}]_{\t_{2}}\times\nn
	&&\hspace{70pt} \times\,
	Z_{1}^{\,\t_{1}}\,Z_{2}^{\,\t_{2}}\,Z_{3}^{\,\t_{3}}\,
	Y_{1}^{\s_{1}}\,Y_{2}^{\s_{2}-\t_{3}}\,Y_{3}^{\s_{3}-\t_{2}}\,
	[G_{1}(Y,Z)]^{\u}+\mathcal O\Big(\big(\tfrac{\hat\delta}L\big)^{|n|+1}\Big)\,.
\ea
In this form, it is straightforward to check that each leading term above can be reproduced by the functions \eqref{Q Z pm}.
As an example, let us assume \mt{\bar\mu_{1}\ge\mu_{1}}\,. In this case,
the function \eqref{Pn(x)} in $B^{[n+\frac{\bar\mu_{1}-\mu_{1}}2]}_{P2}(\bar\mu_{1},\mu_{1})$
has to satisfy the condition \mt{\s_{3}\ge (\bar\mu_{1}-\mu_{1})/2+n}\,. As a consequence,
 the minimum power of $Y_{3}$ is
\mt{(\bar\mu_{1}-\mu_{1})/2+\t_{3}}\,, coinciding with the one in \eqref{Q Z pm}.
All in all, we have
\be
	{\rm Span}\Big[{\bm B}_{P1}(\bar\mu_{1},\mu_{1})\,\cup\,
	{\bm B}_{P2}(\bar\mu_{1},\mu_{1})\Big]
	={\rm Span}\Big[ {\bm B}_{\tilde Q}(\bar\mu_{1},\mu_{1})
	\,\cup\,{\bm B}_{Q}(\bar\mu_{1},\mu_{1})\Big]\,,
\ee
where
\be\label{BQQQ}
	{\bm B}_{Q}(\bar\mu_{1},\mu_{1}):=
	\bigcup_{\t_2+\t_3\leq\m_1}\!\!\!\!B^{[\t_2,\t_3]}_{Q}(\bar\mu_{1},\mu_{1})\,.
\ee
Let us mention that, though highly redundant, the basis \eqref{BQQQ} proves very convenient in the study of the solution space associated to more than one PDE.

Before moving to the next Section,
let us summarize the general solutions to eq.~\eqref{pmdiff}
in the $\tilde Q$ and $Q$ basis.

\begin{framed}
\noindent\
\underline{\textbf{Solutions to one equation}} \bigskip \\
For arbitrary \mt{\mu_2-\mu_{3}}\,:
\be\label{H sol}
	\tilde C =
	\sum_{\s_{1}=0}^{\m_{1}}Y_{1}^{\,\s_{1}}\,
	\tilde K^{\s_{1}}(Y_{2},Y_{3},Z_{1},\tilde H_{2},\tilde H_{3})\,.
\ee
For \mt{\mu_{1}+\mu_{2}-\mu_{3}\in 2\,\mathbb Z}\,, one also has
\ba\label{shift sol}
	C \eq \sum_{(\t_{1},\t_{2},\t_{3})\in \mathscr L_{1}}
	Z_{1}^{\,\t_{1}}\,Z_{2}^{\,\t_{2}}\,Z^{\,\t_{3}}_{3}\
	Y_{2}^{R\left(\t_{2}+\frac{\mu_{2}-\mu_{3}-\mu_{1}}2\right)}\,
	Y_{3}^{R\left(\t_{3}+\frac{\mu_{3}-\mu_{1}-\mu_{2}}2\right)}\times \nn
	&& \hspace{60pt} \times\,
	e^{-\frac{\hat\delta}L\,\cD}\,K^{\t_{1}\t_{2}\t_{3}}(Y,G)\,\Big|_{\sst G=G(Y,Z)},
\ea
where
\be\label{Li}
	\mathscr L_{i}:=\big\{\, (\t_{1},\t_{2},\t_{3})\in \mathbb N^{3}\ \big|\
	\t_{i+1}+\t_{i-1}\le \mu_{i}\,\big\}\,.
\ee

\end{framed}

\section{Intersection of the solution spaces}
\label{sec: intersection}

Let us now consider general interactions involving more than one (P)M field.
In these cases, depending on their number, one has to consider the intersections of the solution spaces of the corresponding PDEs. In the following, we carry out this analysis
for the $\tilde Q$ and $Q$ solutions separately.
\subsection{$\tilde Q$ solutions}

When two (P)M fields (say \mt{i=1, 2}) are present,
one has to take the intersection between the coupling \eqref{H sol} and
its cyclic permutation:
\be
	\sum_{\s_{1}=0}^{\m_{1}}
	Y_{1}^{\,\s_{1}}\,\tilde K_{1}^{\s_{1}}
	(Y_{2},Y_{3},Z_{1},\tilde H_{2},\tilde H_{3})\,,
	\qquad
	\sum_{\s_{2}=0}^{\m_{2}}
	Y_{2}^{\,\s_{2}}\,\tilde K_{2}^{\s_{2}}
	(Y_{3},Y_{1},Z_{2},\tilde H_{3},\tilde H_{1})\,,
\ee
that is
\be\label{2coupl}
	\tilde C =
	\sum_{\s_{1}=0}^{\mu_{1}}\sum_{\s_{2}=0}^{\m_{2}}\,
	Y_{1}^{\,\s_{1}}\,Y_{2}^{\,\s_{2}}\,
	\tilde K^{\s_{1}\s_{2}}(Y_{3},\tilde H_{1},\tilde H_{2},\tilde H_{3})\,.
\ee
When all three interacting fields are (P)M, one has to intersect the couplings \eqref{2coupl}
with
\be
\sum_{\s_{3}=0}^{\m_{3}}
	Y_{3}^{\,\s_{3}}\,\tilde K_{3}^{\s_{3}}
	(Y_{1},Y_{2},Z_{3},\tilde H_{1},\tilde H_{2})\,,
\ee
leading to
\be
	\tilde C =
	\sum_{\s_{1}=0}^{\mu_{1}}\sum_{\s_{2}=0}^{\m_{2}}\sum_{\s_{3}=0}^{\m_{3}}\,
	Y_{1}^{\,\s_{1}}\,Y_{2}^{\,\s_{2}}\,Y_{3}^{\,\s_{3}}\,
	\tilde K^{\s_{1}\s_{2}\s_{3}}(\tilde H_{1},\tilde H_{2},\tilde H_{3})\,.
\ee

\subsection{$Q$ solutions}

For the analysis of the intersections among
the $Q$ solutions \eqref{shift sol},
it is convenient to start with
the leading terms $C^{\sst (0)}$
of the solutions to the entire system of PDEs.
Plugging the expansion \eqref{1/L exp} into eq.~\eqref{pmdiff}, one discovers
that $C^{\sst (0)}$ satisfies the following relatively simple equations:
\be\label{leading diff}
\left(Y_{i+1}\,\partial_{Z_{i-1}}-Y_{i-1}\,\partial_{Z_{i+1}}\right)^{\mu_i+1}\,
C^{\sst(0)}(Y,Z)=0\,,
\ee
which can be solved
in terms of an arbitrary function $K^{\t_{1}\t_{2}\t_{3}}(Y,G)$ as
\be\label{leading form}
	C^{\sst (0)}(Y,Z)=\sum_{(\t_{1},\t_{2},\t_{3})\in \mathscr L}
	Z_{1}^{\,\t_{1}}\,Z_{2}^{\,\t_{2}}\,Z_{3}^{\,\t_{3}}\,
	K^{\t_{1}\t_{2}\t_{3}}\big(Y,G(Y,Z)\big)\,.
\ee
Here, the lattice $\mathscr L$ is given by the intersection of
the lattices $\mathscr L_{i}$'s \eqref{Li} associated with the $i$-th PM fields.
In the following, we  derive the general solutions to the system of eqs.~\eqref{pmdiff}
starting from the leading terms $C^{\sst (0)}$ of eq.~\eqref{leading form}\,.
For that, we distinguish between the cases where two or
three (P)M fields are involved.

When two (P)M fields  (say \mt{i=1,2}) are present,
 one has to solve the system:
\ba
	&\cL_{1}(\mu_{3}-\mu_{2}-\mu_{1})\,
	\cL_{1}(\mu_{3}-\mu_{2}-\mu_{1}+2)\cdots
	\cL_{1}(\mu_{3}-\mu_{2}+\mu_{1})\
	C(Y,Z)=0\,,
	\label{first eq}\\
	&\cL_{2}(\mu_{1}-\mu_{3}-\mu_{2})\,
	\cL_{2}(\mu_{1}-\mu_{3}-\mu_{2}+2)\cdots
	\cL_{2}(\mu_{1}-\mu_{3}+\mu_{2})\
	C(Y,Z)=0\,,
	\label{scnd eq}
\ea
where
the leading terms of the corresponding couplings are given by \eqref{leading form} with
\be
	\mathscr L= \mathscr L_{1}\cap\mathscr L_{2}\,.
\ee
At this point, it is convenient to split the sum over \mt{(\t_1,\t_2,\t_{3})}
into the two  regions:
\be
	 \mathscr L_{1}\cap\mathscr L_{2}\cap \mathscr L_{3}\,,
	 \qquad
	\mathscr L_{1}\cap\mathscr L_{2}\cap \mathscr L_{3}^{\,c}\,,
\ee
where $\mathscr L_{i}^{\,c}$ denotes the complement of the set $\mathscr L_{i}$\,.
When \mt{(\t_1,\t_2,\t_{3})\in\mathscr L_{1}\cap\mathscr L_{2}\cap \mathscr L_{3}}\,, out of the three combinations
\mt{\t_{i}+\tfrac{\mu_{i}-\mu_{i+1}-\mu_{i-1}}2}\,, at most one can be
positive. Hence, without loss of generality, one can assume
\be\label{b23<0}
	\t_{2}+\tfrac{\mu_{2}-\mu_{3}-\mu_{1}}2\le 0\,,\qquad
	\t_{3}+\tfrac{\mu_{3}-\mu_{1}-\mu_{2}}2\le 0\,.
\ee
Then,  depending on the sign of
\mt{\t_{1}+\tfrac{\mu_{1}-\mu_{2}-\mu_{3}}2}\,, there are two subcases.
When the latter is non-positive,
a comparison with eq.~\eqref{shift sol} and its cyclic permutation
directly shows  that the coupling:
\be\label{sol 1}
	C(Y,Z)=Z_{1}^{\,\t_{1}}\,Z_{2}^{\,\t_{2}}\,Z^{\,\t_{3}}_{3}\
	e^{-\frac{\hat\delta}L\,\cD}\,K^{\t_{1}\t_{2}\t_{3}}(Y,G)\,\Big|_{\sst G=G(Y,Z)}\,,
\ee
is a solution to both PDEs (\ref{first eq}\,,\,\ref{scnd eq}).
On the other hand, when
\be\label{b1>0}
	\t_{1}+\tfrac{\mu_{1}-\mu_{2}-\mu_{3}}2>0\,,
\ee
one can consider the following function:
\be\label{sol 2}
	C(Y,Z)=Z_{1}^{\,\t_{1}}\,Z_{2}^{\,\t_{2}}\,Z^{\,\t_{3}}_{3}\
	Y_{1}^{\t_{1}+\frac{\mu_{1}-\mu_{2}-\mu_{3}}2}\,
	e^{-\frac{\hat\delta}L\,\cD}\,K^{\t_{1}\t_{2}\t_{3}}(Y,G)\,\Big|_{\sst G=G(Y,Z)}\,,
\ee
that is a solution to eq.~\eqref{scnd eq}.
Although it is not manifest, the latter solves also eq.~\eqref{first eq}.
One way of proving it consists in pushing the exponential function to the left, ending up with:
\be
	C(Y,Z) =
	\sum_{p,q,r\ge 0}
	Z_{1}^{\,\t_{1}}\,Z_{2}^{\,\bar\t_{2}}\,Z_{3}^{\,\bar\t_{3}}\,
	e^{-\frac{\hat\delta}L\,\mathcal D}\,\bar K^{\t_{1}\bar\t_{2}\bar\t_{3}}
	_{p,q,r}(Y,G)
	\,\Big|_{\sst G=G(Y,Z)}\,,
\ee
where \mt{\bar\t_{2}=\t_{2}+p+r}\,, \mt{\bar\t_{3}=\t_{3}+q+r} and
\ba
	\bar K^{\t_{1}\bar\t_{2}\bar\t_{3}}_{p,q,r}(Y,G)\eq
	\binom{\t_{1}+\frac{\mu_{1}-\mu_{2}-\mu_{3}}2}{p,q,r}\,
	\big(\tfrac{\hat\delta}L\big)^{p+q+r}\times\nn
	&&\times\,
	Y_{1}^{\t_{1}+\frac{\mu_{1}-\mu_{2}-\mu_{3}}2-p-q-r}
	\partial_{Y_{2}}^{\ q}\,\partial_{Y_{3}}^{\ p}\,\partial_{G}^{\ r}\,
	K^{\t_{1}\t_{2}\t_{3}}(Y,G)\,.
\ea
Comparing with \eqref{shift sol}, one can see that
the latter is a solution to the first equation provided
the conditions:
\ba
	&\t_{1}+\bar\t_{2}\le \mu_{3}\,,\qquad
	\bar\t_{2}+\bar\t_{3}\le \mu_{1}\,,\qquad
	\bar\t_{3}+\t_{1}\le \mu_{2}\,,&\nn
	& \bar\t_{2}+\tfrac{\mu_{2}-\mu_{3}-\mu_{1}}2\le 0\,,\qquad
	\bar\t_{3}+\tfrac{\mu_{3}-\mu_{1}-\mu_{2}}2\le 0\,,
\ea
are satisfied. These conditions hold for any
\mt{(\t_{1},\t_{2},\t_{3})\in\mathscr L_{1}\cap\mathscr L_{2}\cap \mathscr L_{3}}
satisfying conditions \eqref{b23<0} and \eqref{b1>0},
and therefore the function \eqref{sol 2} solves eq.~\eqref{first eq}.
Finally, the solutions \eqref{sol 1} and \eqref{sol 2} can be written at once
in a cyclic form as
\ba\label{shift sol IS}
	C \eq \hspace{-10pt}
	\sum_{(\t_{1},\t_{2},\t_{3})\in \mathscr L_{1}\cap\mathscr L_{2}
	\cap \mathscr L_{3}}\hspace{-10pt}
	Z_{1}^{\,\t_{1}}\,Z_{2}^{\,\t_{2}}\,Z^{\,\t_{3}}_{3}\
	Y_{1}^{R\left(\t_{1}+\frac{\mu_{1}-\mu_{2}-\mu_{3}}2\right)}\,
	Y_{2}^{R\left(\t_{2}+\frac{\mu_{2}-\mu_{3}-\mu_{1}}2\right)}\,
	Y_{3}^{R\left(\t_{3}+\frac{\mu_{3}-\mu_{1}-\mu_{2}}2\right)}\times \nn
	&& \hspace{80pt} \times\,
	e^{-\frac{\hat\delta}L\,\cD}\,K^{\t_{1}\t_{2}\t_{3}}(Y,G)\,\Big|_{\sst G=G(Y,Z)}.
\ea
Let us now consider the case
\mt{(\t_{1},\t_{2},\t_{3})\in\mathscr L_{1}\cap\mathscr L_{2}\cap \mathscr L_{3}^{\,c}}\,, where \mt{\mu_{3}<\t_{1}+\t_{2}\le s_{3}}\,. If one requires that the third field be a unitary massive field, then this type of interactions are ruled out in dS (see Figure \ref{fig}).\footnote{
The third field may be PM, but then one has to impose gauge invariance also under its gauge symmetry. This case is considered later.}
On the other hand, in AdS these would correspond to the interactions between two massless and one massive fields with $\mu_3<0$\,, but
we do not find any solution of this type.

Last of all, when all three fields are (P)M, the leading terms of the corresponding couplings are given by \eqref{leading form} with
\be
	\mathscr L= \mathscr L_{1}\cap\mathscr L_{2}\cap \mathscr L_{3}\,.
\ee
This situation is nothing but a subcase of the two (P)M interactions that we have considered before, therefore the solution is given by \eqref{shift sol IS}.

At this point, we have completed the analysis of the (P)M cubic interactions,
and the results are summarized in Section \ref{sec: summary}.

\section{Discussions}
\label{sec: counting}

To conclude, we discuss the implications
of the condition \eqref{rmmcond1} in more details,
pointing out the key differences with respect to the flat-space case.
Moreover, in order to make contact with our previous work \cite{Joung:2012rv},
we provide an example.

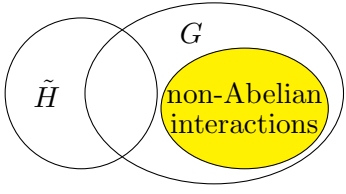
\begin{figure}[h]
\centering
\begin{tikzpicture}
\draw (-0.75,0) ellipse (1cm and 1cm);
\draw (1,0)  ellipse (1.7cm and 1.2cm);
\draw [fill=yellow] (1.4,-0.2) ellipse (1.1cm and 0.8cm);
\node at (-1.2,0) {$\tilde{H}$};
\node at (0.7,0.8) {$G$};
\node at (1.4,-0) {non-Abelian};
\node at (1.4,-0.4) {interactions};
\end{tikzpicture}
\caption{$\tilde{H}$- and $G$-couplings. Non-abelian couplings are a subset of the $G$-couplings that can not be written as $\tilde{H}$-couplings.}
\label{fig2}
\end{figure}

\paragraph{Non-Abelian interactions}
Let us recall that the cubic interactions
which exist regardless of the condition \eqref{rmmcond1}
are given by arbitrary functions of $\tilde H$ ($\tilde Q$ solutions).
Since the latter are trivially gauge invariant---their gauge invariance does
not rely on the on-shell conditions---they do not lead to any deformation of
the gauge transformations.
To iterate, they are all Abelian and of the Born-Infeld type, expressible
in terms of linear curvatures.
When the condition \eqref{rmmcond1} is satisfied,
besides these $\tilde H$-couplings,
supplemental $G$-couplings ($Q$ solutions) appear.
However, only a part of them is independent from the $\tilde H$-couplings
and may lead to non-Abelian deformations of the gauge symmetries (see Figure \ref{fig2} for a schematic picture).
Let us stress that there are many PM interactions which do not satisfy the condition
\eqref{rmmcond1}, including notably the PM spin-2 self-interactions
\cite{Zinoviev:2006im,Deser:2012qg}. Those interactions cannot give rise to any non-Abelian deformations of the gauge symmetries.

\paragraph{Interactions of one massless and two massive fields}

Apart from PM interactions, which do not have any flat-space counterpart,
a novel property of (A)dS is present in
the cubic interactions of one massless ($\m_1=0$) and two massive (\mt{\mu_3\geq\m_2}) fields.
According to the condition  \eqref{rmmcond1},
supplemental $G$ solutions appear when \mt{\mu_3-\mu_2\in 2\,\mathbb{Z}_{\sst \geq 0}}\,. Among them,
the non-$\tilde H$-couplings are the ones satisfying conditions \eqref{con pq} and \eqref{BP2}:
\be
	\s_{1}\ge \s_{2}+\s_{3}+1\,,\qquad \s_{3}\ge \tfrac{\mu_{3}-\m_{2}}2\,,
\ee
where the $\s_{i}$'s are related to the spins $s_{i}$'s as
\be
	s_{1}=\s_{1}+\u\,,\qquad
	s_{2}=\s_{2}+\u+\t_{1}\,,\qquad
	s_{3}=\s_{3}+\u+\t_{1}\,,
\ee
with $\u$ and $\t_{1}$ being the powers of $G_{1}$ and $Z_{1}$ respectively
(see \eqref{P series basis}).
This type of interactions is related to non-trivial Noether currents.
In particular, the electromagnetic (\mt{1\!-\!s\!-\!s}) and the gravitational
(\mt{2\!-\!s\!-\!s}) minimal couplings correspond to \mt{\m_2=\m_3}\,, as in flat space.
On the other hand, there are many interactions
that satisfy the above conditions
for positive even integers \mt{\mu_{3}-\mu_{2}}\,.
For instance, when \mt{\mu_3-\mu_2=2}\,,
one can have \mt{2\!-\!s\!-\!(s+1)} interactions
involving non-trivial Noether currents made of fields with different masses.
This is a novelty with respect to flat space, where
the interactions leading to non-trivial conserved currents
are only available when the two masses are \emph{equal}: \mt{\mu_{2}=\mu_{3}}\,.
This observation suggests that, in (A)dS,
(HS) multiplets may involve not only fields with different spins, but
also with different masses.
Let us also mention that the minimal-like couplings \mt{s_{1}\!-\!s\!-\!s}
require \mt{\m_{3}-\m_{2}\le s_{1}-1}\,, so that in this case interactions
involving fields with different masses
are available only for $s_{1}\ge 3$\,.

\paragraph{Example: \mt{4\!-\!4\!-\!2} interactions}

For concreteness, starting from the general solutions provided in this paper,
we show how to recover the example of  \mt{4\!-\!4\!-\!2} interactions obtained in \cite{Joung:2012rv} by means of a numerical algorithm. The latter
are the cubic interactions  between two spin-$4$ fields at their first PM points (\mt{\mu_1=\mu_2=1}) and a massless spin-$2$ field (\mt{\mu_3=0}).
In this case one can find two $\tilde Q$ solutions \eqref{Qtilde3}:
\be\label{442H+}
\tilde H_1\,\tilde H_2\,\tilde H_3^{\,3}\,,\qquad \tilde H_1\,\tilde H_2\,\tilde H_3^{\,2}\,Y_1\,Y_2\,,
\ee
and six $Q$ solutions \eqref{shift sol IS++}:
\ba\label{442Q}
e^{-\frac{\hat\d}{L}\,\cD}\,Y_1^{\,4}\,Y_2^{\,4}\,Y_3^{\,2}\,,\qquad
&e^{-\frac{\hat\d}{L}\,\cD}\,Y_1^{\,3}\,Y_2^{\,3}\,Y_3\, G\,,\qquad
&e^{-\frac{\hat\d}{L}\,\cD}\,Y_1^{\,2}\,Y_2^{\,2}\, G^2\,,\nn
Z_3\,e^{-\frac{\hat\d}{L}\,\cD}\,Y_1^{\,3}\,Y_2^{\,3}\,Y_3^{\,2}\,,\qquad
&Z_3\,e^{-\frac{\hat\d}{L}\,\cD}\,Y_1^{\,2}\,Y_2^{\,2}\,Y_3\, G\,,\qquad
&Z_3\,e^{-\frac{\hat\d}{L}\,\cD}\,Y_1\,Y_2\, G^2\,.
\ea
However, two of the above $Q$ couplings---the first and the fourth---
can be expressed in terms of the $\tilde Q$ couplings \eqref{442H+}
and the remaining $Q$ couplings.
Hence, one is left with six independent solutions (two $\tilde H$-couplings
and four (non-$\tilde H$) $G$-couplings) in complete agreement with the result of \cite{Joung:2012rv}. More explicitly, for instance,
the lowest-derivative interaction
\mt{Z_3\,e^{-\frac{\hat\d}{L}\,\cD}\,Y_1\,Y_2\, G^2} gives
\be
Z_3\,Y_1\,Y_2\,G^2-\tfrac{\hat{\delta}}L\,Z_3^2\left[G^2+2\,(Y_1\,Z_1+Y_2\,Z_2)\,G+2\,Y_1\,Y_2\,Z_1\,Z_2\right]
+4\,\big(\tfrac{\hat{\delta}}L\big)^2 Z_1\,Z_2\,Z_3^{\,3}\,,
\ee
reproducing the vertex $C_{6}$ in eq.~(3.40) of \cite{Joung:2012rv}.

\acknowledgments{
We are grateful to
A. Sagnotti and  E. Skvortsov for discussions.
The present research was supported in part
by Scuola Normale Superiore,
by INFN, and by the MIUR-PRIN contract 2009-KHZKRX.
The work of MT was also supported  by the Max Planck Institute for Gravitational Physics (Albert Einstein Institute).
}

\appendix

\section{Recurrence relation}
\label{sec: series}

In this Appendix, we provide some details on the derivation of the solutions to the massless eq.~\eqref{massless} and to the PM eq.~\eqref{pmless}.

\subsection{Massless equation}

As explained in Section~\ref{sec: 1 eq},
plugging the expansion:
\be
\label{fullsol1e2+}
C_{\s_{1}}(Y,Z)=\sum_{k=0}^\infty\,
C^{\sst (k)}_{\s_{1}}(Y_{2},Y_{3},Z)\ (-\tfrac{\hat{\delta}}{L}\,\partial_{Y_1})^k\,Y_{1}^{\,\s_{1}}\,,
\ee
into the equation:
\be
\label{massless+}
\Big[\,Y_2\,\partial_{Z_3}-Y_3\,\partial_{Z_2}
+\tfrac{\hat\delta}{L}\left(Y_2\,\partial_{Y_2}-Y_3\,\partial_{Y_3}
+\tfrac{\bar \mu_{1}}2\right)\,\partial_{Y_{1}}\Big]\,C(Y,Z)=0\,,
\ee
one ends up with
a differential recurrence relation for
$C^{\sst (k)}_{\s_{1}}$\,:
\be
\label{reccrelCnn+}
\left(Y_2\,\partial_{Z_3}-Y_3\,\partial_{Z_2}\right)\,C^{\sst (k)}_{\s_{1}}(Y_{2},Y_{3},Z)
=\left(Y_2\,\partial_{Y_2}-Y_3\,\partial_{Y_3}+
\tfrac12\,\bar \mu_{1}\right)C^{\sst (k-1)}_{\s_{1}}(Y_{2},Y_{3},Z)\,,
\ee
where \mt{C^{\sst (-1)}_{\s_{1}}=0}\,.
The latter can be solved iteratively starting from $k=0$\,:
\ba
\label{C0lead+}
C^{\sst (0)}_{\s_{1}}(Y_{2},Y_{3},Z) \eq K\big(Y_2,Y_{3},Z_{1},G_{1}(Y,Z)\big)\nn
	\eq \sum_{\s_{2},\s_{3},\t_{1}, \u}\,K^{\t_{1}}_{\s_{1}\s_{2}\s_{3}\u}\,Z_{1}^{\,\t_{1}}\,
	Y_{2}^{\,\s_{2}}\,Y_{3}^{\,\s_{3}}\,[G_{1}(Y,Z)]^{\u}\,,
\ea
where \mt{G_1(Y,Z):=Y_2\,Z_2+Y_3\,Z_3}\,.
For this purpose, one can consider the following ansatz:
\be
	C_{\s_{1}}(Y,Z)= \sum_{\s_{2},\s_{3},\t_{1},\u}\,K^{\t_{1}}_{\s_{1}\s_{2}\s_{3}\u}\,
	P^{\t_{1}}_{\s_{1}\s_{2}\s_{3}\u}(\bar\mu_{1};Y,Z)\,,
\ee
with
\be
\label{ansatzCk}
P^{{\sst (k)}\t_{1}}_{\s_{1}\s_{2}\s_{3}\u}(\bar\mu_{1};Y_{2},Y_{3},Z)=\sum_{\ell=0}^k \,c_{k,\ell}\,(Z_3\,\partial_{Y_2})^{k-\ell}\,(Z_2\,\partial_{Y_3})^{\ell}\,
Y_{2}^{\,\s_{2}}\,Y_{3}^{\,\s_{3}}\,[G_{1}(Y,Z)]^{\u}\,,
\ee
and turn eq.~\eqref{reccrelCnn+} into a recurrence relation for \mt{c_{k,\ell}}\,:
\ba
&&
(k-\ell)\,c_{k,\ell}+(\ell+1)\,c_{k,\ell+1}=c_{k-1,\ell}\,,\nn
&&
(k-\ell)\left(\s_{3}-\ell\right)c_{k,\ell}-(\ell+1)\left(\s_{2}-k+\ell+1\right)\,c_{k,\ell+1}
=\tfrac{\bar\mu_{1}}2\,c_{k-1,\ell}\,.
\ea
The latter can be straightforwardly solved as
\be
\label{solreccrelcnk+}
c_{k,\ell}=\frac{1}{(k-\ell)!\,\ell!}\,\frac{\left[\s_{2}+\tfrac{\bar\mu_{{1}}}2\right]_{k-\ell}\,\left[\s_{3}-\tfrac{\bar\mu_{{1}}}2\right]_{\ell}}{\left[\s_{2}+\s_{3}\right]_{k}}\,.
\ee
Plugging the coefficients \eqref{solreccrelcnk+} into eq.~\eqref{ansatzCk}
and summing over $k$\,, one gets eq.~\eqref{P series basis}.

To conclude this part of the Appendix, let us provide an identity
involving the function $P^{\t_{1}}_{\s_{1}\s_{2}\s_{3}n}$\,:
\ba\label{P lin comb}
	&&
	\sum_{k=0}^{n}\,(-1)^{k}\,\binom{n}{k}\,
	P^{\t_{1}}_{\s_{1}\s_{2}\s_{3}n}(x-n+2k)\nn
	&&
	=\,\big(\tfrac{\hat\delta}L\big)^{n}\,\left(Z_{2}\,\partial_{Y_{3}}-Z_{3}\,\partial_{Y_{2}}\right)^{n}
	\times\nn
	&&\quad \times\,
	\Bigg[\sum_{p,q\ge 0}\,\frac{\left[\sigma_2+\frac{x}2\right]_{p}
	\left[\sigma_3-\frac{x}2\right]_q}{\left[\sigma_2+\sigma_3\right]_{p+q+n}}\,
	\frac{\big(-\frac{\hat\delta}L\,Z_3\,\partial_{Y_{1}}\,\partial_{Y_2}\big)^{p}}{p!}\,
	\frac{\big(-\frac{\hat\delta}L\,Z_2\,\partial_{Y_3}\,\partial_{Y_{1}}\big)^{q}}{q!}\,
	\Bigg]\times \nn
	&&\quad \times\phantom{\big|}
	Z_{1}^{\,\t_{1}}\,Y_{1}^{\,\s_{1}}\,Y_{2}^{\,\s_{2}}\,Y_{3}^{\,\s_{3}}\,
	[G_{1}(Y,Z)]^{\u}\,.
\ea

\subsection{PM equation}

The first step consists in recasting the higher-order PDE:
\be\label{pmless+}
	\cL_{1}(\bar \mu_{1}-\mu_{1})\,\cL_{1}(\bar \mu_{1}-\mu_{1}+2)
	\cdots \cL_{1}(\bar \mu_{1}+\mu_{1})\,C(\bar\mu_{1},\mu_{1};Y,Z)=0\,,
\ee
into the following system of equations:
\ba
\begin{cases}
\label{systempm}
\cL_{1}(\bar \mu_{1}-\mu_{1})\,\bar C_{\sst[\mu_{1}]}=0\\
\cL_{1}(\bar \mu_{1}-\mu_{1}+2)\,\bar{C}_{\sst[\mu_{1}-1]}=\bar C_{\sst[\mu_{1}]} \\
\hspace{40pt}\vdots\\
\cL_{1}(\bar \mu_{1}+\mu_{1})\,C=\bar{C}_{\sst[1]}
\end{cases}\hspace{-12pt}.
\ea
Afterwards, one expands the $C$ and $\bar{C}_{\sst [n]}$'s as in eq.~\eqref{fullsol1e2+}\,,
so that the system \eqref{systempm} translates into a set of differential recurrence relations for $C^{\sst (k)}_{\s_{1}}$ and
$\bar C^{\sst (k)}_{{\sst[n]}\s_{1}}$\,:
\ba
\begin{cases}
\label{systempm2}
\left(Y_2\,\partial_{Z_3}-Y_3\,\partial_{Z_2}\right)\,\bar C^{\sst (k)}_{{\sst[\mu_{1}]}\s_{1}}
=\left(Y_2\,\partial_{Y_2}-Y_3\,\partial_{Y_3}+
\tfrac{\bar \mu_{1}-\mu_1}{2}\right)\bar C^{\sst (k-1)}_{{\sst[\m_{1}]}\s_{1}}\\
\left(Y_2\,\partial_{Z_3}-Y_3\,\partial_{Z_2}\right)\,\bar{C}^{\sst (k)}_{{\sst[\mu_{1}-1]}\s_{1}}
=\left(Y_2\,\partial_{Y_2}-Y_3\,\partial_{Y_3}+
\tfrac{\bar \mu_{1}-\mu_1+2}{2}\right)\bar{C}^{\sst (k-1)}_{{\sst[\m_{1}-1]}\s_{1}}+\bar C^{\sst (k)}_{{\sst[\m_{1}]}\s_{1}}\\
\hspace{120pt}\vdots
\\
\left(Y_2\,\partial_{Z_3}-Y_3\,\partial_{Z_2}\right)\,{C}^{\sst (k)}_{\s_{1}}
=\left(Y_2\,\partial_{Y_2}-Y_3\,\partial_{Y_3}+
\tfrac{\bar \mu_{1}+\mu_1}{2}\right){C}^{\sst (k-1)}_{\s_{1}}
+\bar{C}^{\sst (k)}_{{\sst[1]}\s_{1}}
\end{cases}\hspace{-12pt},
\ea
where \mt{\bar{C}^{\sst (-1)}_{{\sst[n]}\s_{1}}=0}\,. The solutions ${C}^{\sst (k)}_{\s_{1}}$
to the above system can be written as
\be\label{Cksigma}
	{C}^{\sst (k)}_{\s_{1}}(\bar\mu_{1},\mu_{1};Y,Z)
	=\sum_{n=0}^{\mu_{1}}
	\sum_{\ell=0}^n \frac{(-1)^{n-\ell}}{(n-\ell)!\,\ell!}\,
	C^{\sst (k+n)}_{{\sst [n]}\s_{1}}(\bar\mu_{1}+\mu_{1}-2\,\ell,0;Y,Z)\,,
\ee
where the $C_{\sst [n]\s_{1}}^{\sst (k)}$'s are the expansion coefficients of $C_{\sst [n]\s_{1}}$\,, satisfying
\be
	\mathcal L_{1}(x)\,C_{\sst [n]\s_{1}}(x;Y,Z)=0\,.
\ee
Let us stress once again that,
since the $C_{\sst [n]\s_{1}}$'s
may involve singular terms,
 the expression \eqref{Cksigma}
is only a formal solution.
Hence, for any (formal) solution $C_{\s_{1}}$
to the PM equation, there exist
(formal) solutions $C_{{\sst [n]}\s_{1}+n}$
to the massless equation such that
\be
	C_{\s_{1}}(\bar\mu_{1},\mu_{1};Y,Z)
	=\sum_{n=0}^{\mu_{1}}
	\sum_{\ell=0}^{n}\,(-1)^{\ell}\,\binom{n}{\ell}\,
	C_{{\sst [n]}\,\s_{1}+n}(\bar\mu_{1}+\mu_{1}-2\,\ell;Y,Z)\,.
\ee

\section{Shift solutions}
\label{sec: shift}

In this Appendix, we prove that the function \eqref{Q Z pm} is
a solution to
the PM equation \eqref{pmdiff}.
For that, let us first generalize the identities \eqref{shift id} to
\ba
	&& \cL_{1}(x-n)\,\cdots\,\cL_{1}(x+n-2)\,\cL_{1}(x+n)\,Z_{2}\nn
	&&=\,\big[ Z_{2}\,\cL_{1}(x+n)-(n+1)\,Y_{3} \big]\,
	\cL_{1}(x-n)\,\cdots\,\cL_{1}(x+n-2)\,,\nn
	&& \cL_{1}(x-n)\,\cL_{1}(x-n+2)\,\cdots\,\cL_{1}(x+n)\,Z_{3}\nn
	&&=\,\big[ Z_{3}\,\cL_{1}(x-n)+(n+1)\,Y_{2} \big]\,
	\cL_{1}(x-n+2)\,\cdots\,\cL_{1}(x+n)\,.
\ea
One can see that the effect of $Z_{2}$ (or $Z_{3}$)
is to remove the differential operator $\cL_{1}$
with the largest (or the smallest) argument.
Hence, considering generic powers $Z_{2}^{\,\t_{2}}\,Z_{3}^{\,\t_{3}}$ with \mt{\t_{2}+\t_{3}\le \mu_{1}},
any solution to the differential equation:
\be\label{ZZ pm}
	\cL_{1}(\bar\mu_{1}-\mu_{1}+2\,\t_{3})\,
	 \cdots\,\cL_{1}(\bar\mu_{1}+\mu_{1}-2\,\t_{2})\,C^{\t_{2}\t_{3}}(Y,Z)=0\,,
\ee
is also a solution to
\be
	\cL_{1}(\bar\mu_{1}-\mu_{1})\,\cdots\,\cL_{1}(\bar\mu_{1}+\mu_{1})\,
	 Z_{2}^{\,\t_{2}}\,Z_{3}^{\,\t_{3}}\,C^{\t_{2}\t_{3}}(Y,Z)=0\,.
\ee
Moreover, as any solution to a single $\cL_{1}$ equation:
\be
	\cL_{1}(\bar\mu_{1}-\mu_{1}+2\,\t_{3}+2\,n)\,C^{\t_{2}\t_{3}}(Y,Z)=0\,,
\ee
also solves the equation \eqref{ZZ pm} for any
\mt{n=0,1,\ldots,\mu_{1}-\t_{2}-\t_{3}}\,,
one can choose the value of $n$
which minimizes
\mt{|\bar\mu_{1}-\mu_{1}+2\,\t_{3}+2\,n|}\,.
This corresponds to the solution \eqref{Q Z pm}.

\section{H solutions}
\label{sec: H}

In this appendix we show how to turn a generic function of $\tilde H$ into a function
of $Y$ and $Z$ after integrating by parts all the total-derivative terms present in
\eqref{tildeH}.
Let us start considering the integration by parts of a single $\tilde H_i$\,:
\mt{\tilde H_i\,K(Y,Z,\tilde H)\,\simeq\,\hat H_i\,K(Y,Z,\tilde H)}\,.
Here the $\hat H_i$'s are operators defined as
\be
\hat H_i:=Y_{i-1}\,Y_{i+1}-\tfrac{\hat \d}{L}\,N_i\,Z_i\,,
\ee
where
\be
N_i:=Y_{{i+1}}\,\partial_{Y_{{i+1}}}+Y_{{i-1}}\,\partial_{Y_{{i-1}}}-Y_{i}\,\partial_{Y_{i}}+Z_{{i}}\,\partial_{Z_{{i}}}+\tfrac{\m_{{i}}-\m_{{i+1}}-\m_{{i-1}}}2\,,
\ee
and $\simeq$ means equality under the integral sign and modulo TT. The above identity suffices to integrate by parts any function of $\tilde H$ as
\be\label{K inte bp}
K(Y,Z,\tilde H)=\!\!\sum_{h_{1},h_{2},h_{3}}\tilde H_1^{h_1}\,\tilde H_2^{h_2}\,\tilde H_3^{h_3}\,K_{h_{1}h_{2}h_{3}}(Y,Z)
\,\simeq \!\!\sum_{h_{1},h_{2},h_{3}}
\hat H_1^{h_1}\,\hat H_2^{h_2}\,\hat H_3^{h_3}\,K_{h_{1}h_{2}h_{3}}(Y,Z)\,.
\ee
The operators $\hat H_i$ and $\hat H_j$ commutes when $i\neq j$\,,
while $\hat H_i^{\,h_{i}}$ gives
\be
\hat H_i^{h_i}=\sum_{k=0}^{h_{i}}\,
\binom{h_{i}}{k}\,[N_i+h_i]_{k}\,
\big(-\tfrac{\hat\d}{L}\,Z_i\big)^{k}\,
(Y_{i+1}\,Y_{i-1})^{h_{i}-k}\,.
\ee
Using the above identity, one can recast eq.~\eqref{K inte bp}
into a compact form as
\be
	K(Y,Z,\tilde H)\simeq
	\left[\,\prod_{i=1}^{3}\left(1-\tfrac{\hat \d}L\,Z_i\,\partial_{H_i}\right)^{\!N_{i}+H_{i}\partial_{H_{i}}}\right]
	K(Y,Z,H)\,\Big|_{H_i=Y_{i-1}Y_{i+1}}\,.\label{H by parts}
\ee
In the following, we find a set of functions of $\tilde H$
which explicitly give all possible leading terms.

\paragraph{One massless\,--\,two massive case}

We aim at finding the functions:
\be K(Y_2,Y_3,\tilde H_2,\tilde H_3,Z_1)\,, \ee
that, after integration by parts, have leading terms involving \mt{G_1^{\,n+1}}\,.
Starting from the identity \eqref{H by parts} and requiring that the first
\mt{n+1} leading terms cancel, one gets
\ba\label{tilde Q massless}
&&Z_1^{\,\t_1}\,Y_2^{\,\s_2}\,Y_3^{\,\s_3}\,\tilde H_2^{\,h_{2}}\,	
\tilde H_3^{\,h_3}\times\nn
&&
\times\,(Y_2\,\tilde H_2-Y_3\,\tilde H_3)\sum_{k=0}^{n}\,\binom{n}{k}\,
(x-n)_{k}\,[x+n]_{n-k}\,(Y_2\,\tilde H_2)^k\,(-Y_3\,\tilde H_3)^{n-k}\nn
&&\simeq\,\big(\tfrac{\hat\delta}L\big)^{n+1}\,[x+n]_{2n+1}\,Z_1^{\,\t_1}\,
\hat H_2^{\,h_{2}}\,\hat H_3^{\,h_3}\,
Y_2^{\,\s_2}\,Y_3^{\,\s_3}\,[G_{1}(Y,Z)]^{n+1}\nn
&&=\,\big(\tfrac{\hat\delta}L\big)^{n+1}\,[x+n]_{2n+1}\,Z_1^{\,\t_1}\,
Y_{1}^{\,h_{2}+h_{3}}\,Y_{2}^{\,\s_{2}+h_{3}}\,Y_{3}^{\,\s_{3}+h_{2}}\,[G_{1}(Y,Z)]^{n+1}
+\mathcal O(\hat\delta^{n+2})\,.
\ea
Here, \mt{[a]_n:=a(a-1)\cdots(a-n+1)} and \mt{(a)_n:=a(a+1)\cdots(a+n-1)} are the descending and ascending Pochhammer symbols respectively, while
\mt{x=(\m_3-\m_2)/2+\s_{2}-\s_{3}}\,.
Let us notice that the results come with factors
which vanish when \mt{x\in\{-n,-n+1,\ldots,n\}}\,.
Hence,  one has to normalize the corresponding leading terms
in order to have non-vanishing couplings also for these values of $x$\,.

\paragraph{Two massless\,--\,one massive case}

In this case, we look for the functions:
\be
	K(Y_3,\tilde H_1,\tilde H_2,\tilde H_3)\,,
\ee
which give rise to leading terms proportional to $G^{n+1}$\,, and we get
\ba\label{tilde Q 2massless}
&& Y_3^{\s_3}\,\tilde H_1^{\,h_{1}}\,\tilde H_2^{\,h_{2}}\,\tilde H_3^{h_3}\times\nn
&& \times\,(\tilde H_1\,\tilde H_2-Y^{\,2}_3\,\tilde H_3)
\sum_{k=0}^{n}\binom{n}{k}\,(x-2n-1)_{k}\,[x-1]_{n-k}\,
(\tilde H_1\,\tilde H_2)^k\,(-Y^{\,2}_3\,\tilde H_3)^{n-k}\nn
&&\simeq\,\big(\tfrac{\hat\delta}L\big)^{n+1}\, [x-1]_{2n+1}\,
Y_{1}^{\,h_{2}+h_{3}}\,Y_{2}^{\,h_{3}+h_{1}}\,Y_{3}^{\,\s_{3}+h_{1}+h_{2}+n+1}\,
[G(Y,Z)]^{n+1}
+\,\mathcal O(\hat\delta^{n+2})\,,\qquad
\ea
with \mt{x=\mu_{3}/2-\s_{3}}\,.

\paragraph{One PM\,--\,two massive case}

In the PM case, a generic $\tilde Q$ coupling is of the form:
\be\label{PM sig eq}
\sum_{\s_{1}=0}^{\m_1}Y_1^{\s_{1}}\,\tilde K^{\s_{1}}(Y_2,Y_3,\tilde H_2,\tilde H_3,Z_1)\,.
\ee
On the other hand, the leading terms of  the corresponding $P1$ solutions,
containing a given number $n$ of $Z_2$ or $Z_3$ (which
can be obtained by using the identity \eqref{P lin comb}),
are
\be
	(Z_2\partial_{Y_3}-Z_3\partial_{Y_2})^k\,
	G_1^{\,n-k}\,Y_1^{\,\s_1}\,Y_2^{\,\s_2+k}\,Y_3^{\,\s_3+k}\,Z_1^{\,\t_1}\,,
	\qquad k=0,\ldots,\text{min}\{n,\m_1\}\,,\label{leading termsss}
\ee
where the additional inequality
\be
\s_2+\s_3+n\geq\s_1\,,\label{ineq}
\ee
is enforced. In general, it is rather involved to find a combination of $\tilde H$
giving rise to each of the leading terms \eqref{leading termsss}
since in this case the solutions will be of the form
\ba
&& Z_1^{\,\t_1}\,Y_1^{\,\s_1-h}\,Y_2^{\,\s_2-h_3}\,Y_3^{\,\s_3-h+h_3}\,
\tilde H_2^{\,h-h_3}\,\tilde H_3^{\,h_3}\,\cP_n^{(k)}(Y_2\,\tilde H_2,Y_3\,\tilde H_3,
Y_1\,Y_2\,Y_3)\\
&& \simeq\big(\tfrac{\hat\d}{L}\big)^{n}\, Z_1^{\,\t_1}(Z_2\partial_{Y_3}-Z_3\partial_{Y_2})^k\,G_1^{\,n-k}\,Y_1^{\,\s_1}\,Y_2^{\,\s_2+k}\,Y_3^{\,\s_3+k}+\mathcal O(\hat\d^{n+1})\,.
\ea
Here, we have introduced the combination:
\be\label{H comb}
\cP_n^{(k)}(Y_2\,\tilde H_2,Y_3\,\tilde H_3,Y_1\,Y_2\, Y_3):=
\sum_{a+b+c=n} p^{(k)}_{abc}\,(Y_2\,\tilde H_2)^a(Y_3\,\tilde H_3)^b(Y_1\, Y_2\, Y_3)^c\,,
\ee
where the  coefficients $p^{(k)}_{abc}$ have to satisfy a system of linear equations.
Let us notice that  the inequality \eqref{ineq} is automatically satisfied due to
the form of  $\tilde H$\,.
For a given order $n$ in $Z_2$ and $Z_3$,
one can count $n+1$ different polynomials in $Y_2\,Z_2$ and $Y_3\,Z_3$\,.
Hence, the total number of polynomials with at most $n$ powers of $Y_2\,Z_2$ and $Y_3\,Z_3$ is $\tfrac{(n+1)(n+2)}2$\,, matching the number of different coefficients $p^{(k)}_{abc}$ entering the combination \eqref{H comb}. This is equivalent to saying that, allowing enough powers of $Y_1$, any leading term in \eqref{leading termsss} can be reproduced after integration by parts by a suitable choice of the coefficients in eq.~\eqref{H comb}.
In the PM case, one has to  impose also the bounds $c\leq \m_1$ in eq.~\eqref{H comb}
and $k\leq \m_1$ for the leading terms of eq.~\eqref{leading termsss}. Hence, the counting of available leading terms with at most $n$ powers of $Y_2\,Z_2$ and $Y_3\,Z_3$ is
\ba
(\rm{number\ of\ leading\ terms})\eq
\tfrac{(\m_1+1)(\m_1+2)}2+(\m_1+1)(n-\m_1)\nn
\eq\tfrac{1}{2}\,(\m_1+1)(2\,n+2-\m_1)\,,
\ea
while the number of free coefficients in \eqref{H comb} satisfying $c\leq\m_1$ is
\be
\big({\rm number\ of}\ p^{(k)}_{abc}\big)=\sum_{\ell=0}^{\m_1}\,(m-\ell+1)=
\tfrac{1}{2}\,(\m_1+1)(2\,n+2-\m_1)\,.
\ee
Therefore, as the number of solutions matches,
using completeness of the $P1$ solutions when \eqref{rmmcond1} is not satisfied,
one can conclude that the $\tilde Q$ solutions span the same space.

In order to clarify the latter discussion, we provide the explicit example of $n=2$\,.
The combination proportional to $G_1^{\,2}$ is given by
\ba
&& Z_1^{\,\t_1}\,Y_1^{\,\s_1-h}\,Y_2^{\,\s_2-h_3}\,Y_3^{\,\s_3-h+h_3}\,
\tilde H_2^{\,h-h_3}\,\tilde H_3^{\,h_3}\,\Big\{\,[x+y+1]_{3}\,(x-y+1)\,(Y_2\,\tilde H_2)^2\nn
&&\quad
-\,2\,(x+y)_{2}\,[x-y]_{2}\,(Y_2\,\tilde H_2)\,(Y_3\,\tilde H_3)
+(x+y-1)\,(x-y-1)_{3}\,(Y_3\,\tilde H_3)^2\nn
&&\quad-\,4\,y\left[(x+y)_{2}(x-y)\,(Y_2\,\tilde H_2)
-(x+y)\,[x-y]_{2}\,(Y_3\,\tilde H_3)\right](Y_1\,Y_2\,Y_3)\nn
&&\quad+\,2\,y\,(2y-1)\,(x+y+1)\,(x-y-1)\,(Y_1\,Y_2\,Y_3)^2\Big\}\,,
\ea
while the one proportional to $G_1\,(Y_2\,Z_2-Y_3\,Z_3)$ is
\ba
&&Z_1^{\,\t_1}\,Y_1^{\,\s_1-h}\,Y_2^{\,\s_2-h_3}\,Y_3^{\,\s_3-h+h_3}\,
\tilde H_2^{h-h_3}\,\tilde H_3^{\,h_3}\times\nn
&&\times\,\Big\{\,
[x+y+1]_{3}\,(x-y+1)\,(Y_2\,\tilde H_2)^2-
(x+y-1)\,(x-y-1)_{3}\,(Y_3\,\tilde H_3)^2\nn
&&\quad\ \  -\,2\,y\left[
(x+y-1)_{3}\,(x-y)\,(Y_2\,\tilde H_2)
-(x+y)\,(x-y-1)_{3}\,(Y_3\,\tilde H_3)\right](Y_1\,Y_2\,Y_3)\nn
&&\quad\ \ +\,2\,(2y-1)\,x\,(x+y+1)\,(x-y-1)\,(Y_1\,Y_2\,Y_3)^2\Big\}\,.
\ea
Finally, the combination proportional to $(Y_2\,Z_2-Y_3\,Z_3)^2$ reads
\ba
&&Z_1^{\,\t_1}\,Y_1^{\,\s_1-h}\,Y_2^{\,\s_2-h_3}\,Y_3^{\,\s_3-h+h_3}\,
\tilde H_2^{\,h-h_3}\,\tilde H_3^{\,h_3}\,
\Big\{\,[x+y+1]_{3}\,(x-y+1)\,(Y_2\,\tilde H_2)^2\nn
&&\quad+\,2\,(x+y)_{2}\,[x-y]_{2}\,(Y_2\,\tilde H_2)\,(Y_3\,\tilde H_3)
+(x+y-1)\,(x-y-1)_{3}\,(Y_3\,\tilde H_3)^2\nn
&&\quad-\,4\,y\left[(x-1)\,(x+y)_{2}\,(x-y)\,(Y_2\,\tilde H_2)
+(x+1)\,(x+y)\,[x-y]_{2}\,(Y_3\,\tilde H_3)\right](Y_1\,Y_2\,Y_3)\nn
&&\quad+\,2\,(2\,x^2-y)\,(x-y-1)\,(x+y+1)\,(Y_1\,Y_2\,Y_3)^2\Big\}\,.
\ea
Here, $x$ and $y$ are defined as
\be
x=\tfrac{\m_2-\m_3}2+\s_3-\s_2\,,\qquad y=\tfrac{\m_1}2-\s_1\,.
\ee
As one can see from the above example, if no bound on the powers of $Y_1$ is imposed
in \eqref{PM sig eq}, then all possible leading terms can be reproduced.
However, for given values of $y$\,, only a subset of the above solutions
satisfy $c\leq\m_1$. For instance, when \mt{\m_1=0}
one can see that only the leading terms proportional to $G_1^{\,n}$
are leftover.

\section{Flat-space limit}
\label{sec: flat}

In this appendix, we present another argument
for the equivalence between the $\tilde Q$ solutions
and $P1$ solutions.
It relies on the limiting process
where both $L$ and $\mu$ go to infinity in such a way that
their ratio stays finite:
\be\label{mass lim}
	\lim_{L\to\infty}\,\frac{|\mu|}L=M\,.
\ee
Let us notice that, as the supplemental $Q$ and $P2$ solutions have to satisfy
the conditions \eqref{rmmcond1},
they cannot be taken into account in this limit so that effectively we
are restricting the attention to the $\tilde Q$ and $P1$ solutions only.
In the \eqref{mass lim} limit, the PDE \eqref{pmless} reduces to
\be
\label{flatPDE}
\Big[Y_2\,\partial_{Z_3}-Y_3\,\partial_{Z_2}+\tfrac{\hat{\delta}}{2}\left(M_2-M_3\right)\partial_{Y_1}\Big]^{\m_1+1}\, C(Y,Z)=0\,,
\ee
which, for \mt{\m_1=0}\,, corresponds to the usual flat Noether procedure equation for one massless and two massive fields.
However, in general, it does not have a direct physical interpretation.
The PDE \eqref{flatPDE} can be easily solved as
\be
C =\sum_{\s_{1}=0}^{\m_{1}}Y_{1}^{\,\s_{1}}\,
	 K^{\s_{1}}(Y_{2},Y_{3},Z_{1}, H_{2}, H_{3})\,,
\ee
where the $H_i$'s are given by
\ba
\label{flatH}
H_i&=&Y_{i+1}\,Y_{i-1}+\tfrac{\hat{\delta}}2\,\left(M_i-M_{i+1}-M_{i-1}\right)\,Z_i\,,\nn
&=&Y_{i+1}\,Y_{i-1}-\tfrac{1}2\,\partial_X\cdot\left(\partial_{X_i}-\partial_{X_{i+1}}-\partial_{X_{i-1}}\right)\,Z_i\,.
\ea
The (A)dS counterpart of \eqref{flatH} can be obtained by adding proper total-derivative terms to the $Y_{i\pm 1}$'s, ending up with $\tilde H_{i}$ \eqref{tildeH}.
In this way,
keeping the number of independent solutions unchanged,
one recovers the following (A)dS vertices:
\be
C =\sum_{\s_{1}=0}^{\m_{1}}Y_{1}^{\,\s_{1}}\,
	 \tilde K^{\s_{1}}(Y_{2},Y_{3},Z_{1}, \tilde H_{2}, \tilde H_{3})\,,
\ee
in agreement with eq.~\eqref{H sol}.
Notice that, in the limit,
the combinations \eqref{tilde Q massless} and \eqref{tilde Q 2massless},
giving rise to $G_{1}^{n}$- and $G^{n}$-type leading terms,
become proportional to
\be
(Y_2\,H_2-Y_3\,H_3)^n\,,
\qquad
(H_{1}\,H_{2}-Y_{3}^{\,2}\,H_{3})^{n}\,,
\ee
respectively, that are nothing but the corresponding flat combinations.

\bibliographystyle{JHEP}
\bibliography{ref}
\providecommand{\href}[2]{#2}\begingroup\raggedright

\end{document}